\documentclass[a4paper,fleqn]{cas-sc}

\usepackage[authoryear,longnamesfirst]{natbib}
\usepackage[utf8]{inputenc}
  \usepackage{algorithm}
  \usepackage{algpseudocode}

\def\tsc#1{\csdef{#1}{\textsc{\lowercase{#1}}\xspace}}
\tsc{WGM}
\tsc{QE}
\tsc{EP}
\tsc{PMS}
\tsc{BEC}
\tsc{DE}
\begin{document}
\let\WriteBookmarks\relax
\def\floatpagepagefraction{1}
\def\textpagefraction{.001}
\shorttitle{Event-Triggered Adaptive Consensus for Multi-Robot Task Allocation}

\shortauthors{F. Aznar, M. Pujol, A. Díez}

\title [mode = title]{Event-Triggered Adaptive Consensus for Multi-Robot Task Allocation}

\author[1]{Fidel Aznar}[orcid=0000-0003-4521-956X]
\ead{fidel@ua.es}
\author[1]{Mar Pujol}[orcid=0000-0002-8575-0067]
\ead{mar.pujol@ua.es}
\author[1]{\'{A}lvaro D\'{i}ez}[orcid=0009-0004-6036-7733]
\ead{alvaro.diez@ua.es}

\affiliation[1]{organization={Department of Computer Science and Artificial Intelligence, University of Alicante},
                city={Alicante},
                country={Spain}}

\begin{abstract}
Coordinating robotic swarms in dynamic and communication-constrained environments remains a fundamental challenge for collective intelligence. This paper presents a novel framework for event-triggered organization, designed to achieve highly efficient and adaptive task allocation in a heterogeneous robotic swarm. Our approach is based on an adaptive consensus mechanism where communication for task negotiation is initiated only in response to significant events, eliminating unnecessary interactions. Furthermore, the swarm self-regulates its coordination pace based on the level of environmental conflict, and individual agent resilience is managed through a robust execution model based on Behavior Trees. This integrated architecture results in a collective system that is not only effective but also remarkably efficient and adaptive. We validate our framework through extensive simulations, benchmarking its performance against a range of coordination strategies. These include a non-communicating reactive behavior, a simple information-sharing protocol, the baseline Consensus-Based Bundle Algorithm (CBBA), and a periodic CBBA variant integrated within a Behavior Tree architecture. Furthermore, our approach is compared with Clustering-CBBA (C-CBBA), a state-of-the-art algorithm recognized for communication-efficient task management in heterogeneous clusters. Experimental results demonstrate that the proposed method significantly reduces network overhead when compared to communication-heavy strategies. Moreover, it maintains top-tier mission effectiveness regarding the number of tasks completed, showcasing high efficiency and practicality. The framework also exhibits significant resilience to both action execution and permanent agent failures, highlighting the effectiveness of our event-triggered model for designing adaptive and resource-efficient robotic swarms for complex scenarios.
\end{abstract} 


\maketitle

\section{Introduction}
Replicating the remarkable efficiency and scalability of biological swarms remains a central goal in robotics. Swarm intelligence offers a powerful paradigm for tackling complex, large-scale problems such as Search and Rescue (SAR), where decentralized teams of robots can cover vast areas and adapt to dynamic events more effectively than a single entity. However, a critical gap persists between natural systems and their robotic counterparts: sustainable resource management. While biological swarms coordinate through highly efficient local interactions, robotic swarms often struggle in real-world, communication-constrained environments. The fundamental challenge is no longer just about achieving coordination, but about doing so efficiently and robustly, ensuring that limited resources like bandwidth and energy are not wasted on unnecessary communication.

Real-world deployments present inherent complexities that challenge robotic coordination. Many operational environments are intrinsically unstructured, dynamic, and offer only incomplete information \cite{liResilientAdaptiveReplanning2024}. Success in these settings demands rapid adaptation to unforeseen events and effective management of uncertainty. Furthermore, communication—a cornerstone of coordination—is often intermittent and degraded; wireless signals can be attenuated or blocked, resulting in limited bandwidth and packet loss \cite{bravo-arrabalStrengtheningMultiRobotSystems2025, francosRoleOpportunitiesTeamwork2023}. The time-critical nature of many applications, where swift execution directly impacts mission success, adds another layer of pressure for efficient decision-making and action.

At its core, effective swarm behavior relies on solving the Multi-Robot Task Allocation (MRTA) problem. In dynamic scenarios such as logistics, environmental monitoring, or disaster response, where tasks emerge unexpectedly and robot failures are common, this challenge is particularly acute. Historically, decentralized allocation strategies have presented a difficult trade-off. On one hand, traditional consensus-based algorithms provide robust and interpretable coordination but often rely on fixed assumptions of perfect or periodic communication. In realistic, communication-degraded environments, this leads to network saturation and wasted resources, ultimately causing suboptimal performance or mission failure. On the other hand, recent learning-based approaches can generate highly efficient communication policies, yet their "black-box" nature often lacks the interpretability and guarantees required for safety-critical missions. Consequently, a critical gap exists for a framework that merges the communication efficiency of modern techniques with the robustness and predictability of classical consensus, thus facilitating practical and scalable swarm deployment.

To address these challenges, this paper introduces a novel paradigm for adaptive swarm coordination: event-triggered self-organization. This approach moves beyond traditional periodic or reactive methods by establishing a new framework where intelligent collective behavior emerges from asynchronous, strategically-timed coordination. We achieve this by fundamentally re-engineering the interplay between distributed consensus and individual agent execution. Drawing inspiration from diverse control solutions \cite{shibataDeepReinforcementLearning2023, gielisCriticalReviewCommunications2022,alissaDesignImplementationEventtriggered2021}, we introduce a purpose-built consensus mechanism that activates communication only in response to mission-relevant events. This event-driven logic is built upon the robust foundation of the Consensus-Based Bundle Algorithm (CBBA) \cite{han-limchoiConsensusBasedDecentralizedAuctions2009}, but transforms its consensus phase from a fixed, scheduled process into a dynamic, on-demand negotiation. At the agent level, this paradigm is enabled by a modular Behavior Tree (BT) architecture, which empowers individual robots to manage local contingencies and, crucially, identify the significant state changes that trigger collective coordination. 

The resulting architecture is a cohesive system where agents intelligently self-regulate their communication and coordination pace. This approach allows the swarm to achieve a superior balance between mission performance and resource conservation, excelling in the dynamic and unpredictable environments where traditional methods become unreliable. By fundamentally rethinking when and why robots should coordinate, our framework delivers high task completion rates comparable to communication-heavy strategies, but with a significant reduction in network overhead.

The main contributions of this paper can be summarized as:
 \begin{enumerate}
   \item  A novel framework for event-triggered self-organization. This framework enables a robotic swarm to achieve highly efficient and adaptive task allocation in dynamic, resource-constrained environments by intelligently deciding when to communicate.

   \item A model for emergent intelligent collective behavior. We demonstrate how the swarm uses event-triggered consensus and an adaptive coordination pace to self-regulate its communication, balancing mission performance with resource conservation.

   \item Enhanced swarm resilience through modular execution. We provide evidence that integrating Behavior Trees at the agent level provides a robust mechanism for managing local execution failures, which improves individual resilience without requiring immediate global re-coordination.

   \item A superior balance of performance and efficiency. Through extensive quantitative evaluation, we show that our framework matches the task completion rates of communication-heavy strategies while reducing network overhead often by an order of magnitude.
\end{enumerate}

The remainder of this paper is organized as follows: Section \ref{section:StateArt} reviews the state of the art in CBBA and the use of BTs in multi-robot systems. Section \ref{section:TASK} describes the Task Allocation Problem. Section \ref{section:CBBA-ETC}  details the architecture and methodology of the proposed CBBA-ETC system. Section \ref{sections:Experimental} describes the experimental setup, compared algorithms, and evaluation metrics. Section \ref{section:Results} presents and discusses the experimental results. Finally, Section \ref{section:Conclusions} concludes the paper and outlines directions for future work.
 
\section{State of the Art\label{section:StateArt}}

\subsection{Decentralized Task Allocation: The Consensus-Based Bundle Algorithm (CBBA)}

The Consensus-Based Bundle Algorithm (CBBA) is a cornerstone of decentralized multi-robot task allocation (MRTA) due to its robustness and scalability. It operates through two main iterative phases: bundle building and consensus for conflict resolution \cite{han-limchoiConsensusBasedDecentralizedAuctions2009}. In the bundle building phase, each robot greedily constructs an ordered sequence of tasks, its "bundle", based on individual scoring metrics. Subsequently, during the consensus phase, agents communicate their task bundles and associated bids to neighbors. Through these local interactions, they iteratively resolve conflicts, typically by allowing the agent with the highest bid to win the contested task. The inherent decentralization of CBBA, requiring no central coordinator, makes it resilient to single-point failures and highly scalable.

However, applying CBBA in realistic scenarios presents significant challenges. A primary difficulty lies in managing dynamic task allocation, where tasks can appear or change properties mid-mission. While variants like CBBA with Partial Replanning (CBBA-PR) offer a mechanism for this purpose, the speed of reconvergence is critical for performance \cite{jangSPACEPythonbasedSimulator2024}. The need for efficient dynamic allocation is also highlighted by other decentralized algorithms like Dec-MRTA, which focuses on time-critical tasks in disaster response \cite{ghassemiDecentralizedDynamicTask2019}. Crucially, these dynamic scenarios intensify a core limitation of CBBA: the communication overhead in its consensus phase can become a bottleneck in large-scale systems or bandwidth-limited environments.

In response to these challenges, numerous extensions have been proposed, though they largely focus on optimizing how consensus is reached or how replanning is executed, rather than when communication is fundamentally necessary. For instance, Asynchronous CBBA (A-CBBA) \cite{johnsonImprovingEfficiencyDecentralized2010} addresses computational heterogeneity by allowing agents to operate on their own update cycles, while the aforementioned CBBA-PR enhances replanning efficiency. Similarly, the Consensus-Based Payload Algorithm (CBPA) modifies the core logic to incorporate finite resources \cite{qiuConsensusBasedDynamicTask2024}. Another significant line of research tackles the communication bottleneck by restructuring the network topology itself. The Clustering-CBBA (C-CBBA) approach \cite{dongCommunicationefficientHeterogeneousMultiUAV2025}, for instance, partitions the swarm into geographically-based clusters using the k-means++ algorithm, effectively breaking the large-scale problem into smaller subproblems. Coordination is then managed hierarchically through a two-tiered consensus process: an internal phase within each cluster and an external phase between designated leader agents. This method significantly reduces the number of communication steps required to reach a conflict-free allocation.

While valuable, these methods address problems orthogonal or complementary to the communication bottleneck. Our work, \emph{CBBA-ETC}, focuses on this distinct axis of optimization: intelligently managing network load by deciding the optimal moments to communicate.

\subsection{Managing Complex Behaviors: Behavior Trees in Multi-Robot Systems}

Behavior Trees (BTs) have become a powerful and flexible tool for modeling and controlling autonomous agent behavior, offering key advantages such as modularity, a hierarchical structure, reactivity, and adaptability \cite{ogrenBehaviorTreesRobot2022}. BTs are composed of nodes representing conditions, actions, or control structures (like sequences, fallbacks/selectors, parallels), with each subtree acting as an independent behavioral module. Their hierarchical nature allows complex tasks to be decomposed into simpler subtasks, enhancing readability and maintainability. Executed via periodic "ticks", BTs are inherently reactive, allowing agents to respond dynamically to environmental or internal state changes. This reactivity, coupled with the ability to adapt based on action outcomes (success/failure) and even insert sub-goals at runtime, makes BTs highly suitable for dynamic environments.

In multi-robot systems, BTs are used to specify complex missions that can be dynamically assigned to team members \cite{heppnerBehaviorTreeCapabilities2024}. There are simulators that use BTs for agent controller implementation, facilitating the development of agent-level behaviors that complement MRTA algorithms like CBBA \cite{jangSPACEPythonbasedSimulator2024}. Planning algorithms like MRBTP (Multi-Robot Behavior Tree Planning) have been developed to generate BTs for robot teams with theoretical guarantees \cite{caiMRBTPEfficientMultiRobot2025a}. Frameworks using BTs with Data Distribution Service (DDS) enable asynchronous control of multiple robots and incorporate local BTs for fault recovery \cite{jeongBehaviorTreeBasedTask2022}.  

The integration of BTs with distributed algorithms like CBBA is an area of growing interest. BTs can serve as the execution layer for tasks assigned by CBBA, managing detailed execution and local error handling \cite{ogrenBehaviorTreesRobot2022}. More profoundly, BTs themselves can define the ``capabilities'' or tasks over which robots bid using CBBA, or even be used to communicate complex ``intentions'' or adaptive policies during the consensus process, enriching coordination \cite{heppnerBehaviorTreeCapabilities2024, hullCommunicatingIntentBehaviour2024}. This convergence suggests BTs are becoming integral to synthesizing and verifying multi-robot behavior, especially with advances in automatic BT generation from formal specifications like Linear Temporal Logic (LTL) \cite{neupaneDesigningBehaviorTrees2023}. The combination of expressive BTs, potentially augmented by Large Language Models (LLMs) for behavior generation, with efficient and adaptive task allocation algorithms, promises more sophisticated and adaptable multi-robot collaboration \cite{liLargeLanguageModels2025b}.

\section{Task Allocation Problem\label{section:TASK}}
This section formally defines the multi-robot coordination challenges addressed in this paper. We frame the problem within the context of a decentralized Multi-Robot Task Allocation (MRTA) scenario, characterized by team heterogeneity, a dynamic environment, and operational uncertainty. The subsequent evaluation of our proposed algorithms is conducted within a custom simulation designed to embody these core challenges.

The central problem is to dynamically assign a set of tasks to a team of heterogeneous robots in a decentralized manner to maximize collective performance over a finite mission duration. The system consists of a team of robotic agents and a collection of tasks, each with specific requirements and constraints.

The multi-robot team is composed of $N_R$ agents, which are functionally heterogeneous. This heterogeneity is a critical constraint, meaning that specific robots possess unique capabilities required for certain tasks. The set of robots can be represented as:
$$R = \{r_1, r_2, \dots, r_{N_R}\}$$
Each robot $r_i$ is endowed with a specific capability, $C_{\text{robot}_i}$, from a predefined set of possible capabilities $\{$RED, GREEN, BLUE$\}$. This abstraction represents specialized equipment or functionalities.

The environment contains a set of $N_V$ tasks that emerge dynamically at unpredictable locations. Each task $j$ is characterized by a specific requirement, $C_{\text{wall}_j}$, corresponding to one of the robot capabilities. A robot $r_i$ can only successfully complete task $j$ if its capability matches the task's requirement (i.e., if $C_{\text{robot}_i} = C_{\text{wall}_j}$). The set of tasks is represented as:
$$T = \{t_1, t_2, \dots, t_{N_V}\}$$
The primary objective is to derive a task allocation policy that maximizes the total number of successfully completed tasks within the simulation period. This requires solving a complex assignment problem under several key constraints:
\begin{itemize}
    \item \textbf{Capability Matching:} Tasks must be assigned to robots possessing the corresponding capability.
    \item \textbf{Incomplete Information:} The capability requirement $C_{\text{wall}_j}$ of a task is not known in advance and can only be discovered when a robot moves within close proximity and performs a dedicated "inspection" action. This creates a challenge of exploration and information sharing. This requirement could be relaxed in order to prioritize the need information to perform the bid of consensus-based strategies.
    \item \textbf{Time Constraints:} Tasks are time-sensitive and exist for a finite lifetime. Failure to complete a task within this window results in mission failure for that task, introducing temporal urgency.
    \item \textbf{Decentralized Coordination:} The system must operate without a central coordinator, requiring robots to rely on local perception and peer-to-peer communication to make assignment decisions.
    \item \textbf{Operational Uncertainty:} Robot actions, including movement and task execution, are subject to stochastic failures, requiring robust and resilient strategies.
\end{itemize}

The problem, therefore, involves not only optimizing the final assignment of robots to tasks but also managing the dynamic and uncertain process of task discovery, information gathering, and collaborative decision-making in a communication-constrained environment.

This MRTA can be formulated as an optimization problem aimed at maximizing the cumulative utility obtained from all robot-task assignments. The objective function is:
$$\max \sum_{i=1}^{N_R} \sum_{j=1}^{N_V} U_{ij}x_{ij}$$
where $x_{ij}$ is the binary decision variable, such that $x_{ij}=1$ if robot $i$ is assigned to task $j$, and $x_{ij}=0$ otherwise. The term $U_{ij}$ represents the utility or reward for robot $i$ successfully completing task $j$. In our model, this utility is primarily a function of the distance $d_{ij}$ between the robot and the task, rewarding proximity to encourage efficiency:
$$U_{ij} = \frac{1}{d_{ij} + \epsilon}$$
This objective function is subject to the following constraints:
\begin{itemize}
    \item \textbf{Unique Assignment Constraint:} Each task can be assigned to at most one robot to prevent redundant efforts.
    $$\sum_{i=1}^{N_R} x_{ij} \le 1 \quad \forall j \in T$$
    \item \textbf{Capability Matching Constraint:} An assignment is only valid if the robot's capability $C_{\text{robot}_i}$ matches the task's requirement $C_{\text{wall}_j}$.
    $$x_{ij} = 0 \quad \text{if } C_{\text{robot}_i} \neq C_{\text{wall}_j}, \quad \forall i \in R, j \in T$$
    \item \textbf{Task Capacity Constraint:} Each robot can be assigned a maximum of $L_i$ tasks at any given time. In consensus-based approaches like CBBA, this value is typically set to a small number to maintain focus on high-priority objectives.
    $$\sum_{j=1}^{N_V} x_{ij} \le L_i \quad \forall i \in R$$
    \item \textbf{Time Window Constraint:} Each task $j$ must be completed before its lifetime expires. Let $t_{\text{spawn}_j}$ be the time task $j$ is generated and $t_{\text{complete}_{ij}}$ be the time robot $i$ completes it.
    $$t_{\text{complete}_{ij}} \le t_{\text{spawn}_j} + T_{\text{lifetime}} \quad \forall(i,j) \text{ where } x_{ij}=1$$
    \item \textbf{Binary Decision Variable:} The decision variable must be binary.
    $$x_{ij} \in \{0,1\} \quad \forall i \in R, j \in T$$
\end{itemize}

This mathematical formulation encapsulates the central challenge: solving a decentralized Multi-Robot Task Allocation (MRTA) problem that must effectively manage team heterogeneity, operational uncertainty (such as dynamically emerging tasks and incomplete information), and strict time constraints. This represents the core difficulty in a multitude of real-world applications, including the Search and Rescue (SAR) , logistics, and disaster response scenarios discussed earlier. Successfully addressing this problem requires a solution that is not only capable of finding an optimal assignment but can also do so robustly and efficiently in a dynamic, communication-constrained environment.

To meet these demands, the following section details the architecture and methodology of our proposed system: CBBA-ETC. Our central contribution is a novel framework that fundamentally re-engineers the interplay between distributed consensus and individual agent execution. We will describe the specific mechanisms by which this architecture achieves a superior balance between mission performance and resource conservation.

\section{CBBA-ETC: System Architecture and Methodology\label{section:CBBA-ETC}}

The CBBA-ETC system is designed as an framework for emergent coordination in multi-robot task allocation and execution, specifically in order to address the challenges prevalent in dynamic and resource-constrained operations. This section shows its architecture, detailing the concepts and synergistic interactions of its core components.

The CBBA-ETC architecture is presented as a synergistic framework in which emergent coordination is achieved through several core technologies. At the strategic level, the Consensus-Based Bundle Algorithm (CBBA) is transformed from a static, periodic protocol into a dynamic, on-demand negotiation process. This transformation is principally driven by an Event-Triggered Control (ETC) mechanism that obviates the need for computationally expensive, periodic communication cycles. Under this paradigm, the principles of ETC are embedded within the consensus process, ensuring that communication bandwidth is utilized exclusively for instances of high strategic value, specifically when new information possesses a significant potential to alter the collective task allocation. This fusion results in a system that is inherently adaptive and resource-efficient by design.

The event-triggered strategy is further refined by two key innovations: an adaptive consensus interval and a robust execution model. The coordination frequency of the swarm is not static but is instead dynamically modulated by an adaptive mechanism. This mechanism adjusts the time-based fallback for consensus in response to the perceived level of collective conflict. Consequently, the swarm can self-regulate its coordination frequency, increasing it during periods of high environmental instability and reducing it to conserve resources during stable phases. This strategic layer is supported by the tactical resilience afforded by Behavior Trees (BTs). The BTs function as the foundational execution and state-monitoring framework for each agent. They are responsible not only for managing local contingencies and action failures autonomously but also for identifying the specific state changes, such as task completion, action failure, or the discovery of a new high-priority target, that constitute the "events" for the higher-level ETC logic. In this capacity, the BT provides the essential link between local, tactical execution and global, strategic coordination, enabling the event-triggered paradigm.

\subsection{General Architecture and Decision Cycle}
The agent's high-level decision-making architecture is implemented as a Behavior Tree (BT), which provides a structured, hierarchical, and reactive control flow for managing the complexities of task execution. The logic of a BT is processed from the root node downwards and from left to right in each "tick", which naturally creates a prioritized system of behaviors. This design allows for both deliberate action towards assigned goals and reactive adaptation when goals change or new opportunities arise. The robot's operational cycle is conceptually defined by three primary branches with descending priority:

\begin{enumerate}
    \item \textbf{Target Validation and Action:} As the highest priority, if a task is currently assigned to the robot via CBBA, the BT first validates its continued relevance and the robot's assignment status. If the task is valid, the BT executes a sequence for task completion, which may involve navigation, inspection, and performing the specific rescue action.
    \item \textbf{Task Acquisition (CBBA Process Invocation):} If no valid target is currently assigned, or if an assignment becomes invalidated (e.g., due to losing the task in a consensus round), the BT logic transitions to this lower-priority branch, directing the robot to initiate the full CBBA process: bundle building, conditional consensus, and new target selection.
    \item \textbf{Exploratory Behavior:} As a final fallback, if no task is assigned and the CBBA process does not yield a new assignment, the BT triggers an exploratory behavior (e.g., wandering) to search for new tasks or information, ensuring the robot remains productive.
\end{enumerate}

To provide a formal specification of this control flow, the agent's main operational loop is presented in Algorithm \ref{alg:main_cycle_bt}. This formalization presents the precise interaction between the system's components in each decision cycle.

\begin{algorithm}[H]
\caption{CBBA-ETC Robot Decision Cycle (BT-based Logic)}
\label{alg:main_cycle_bt}
\begin{algorithmic}[1]
\State \textbf{Input:} Robot state $s_i$, Task list $T$, World knowledge $W_i$
\State \textbf{Output:} Executes one action per cycle (tick)

\Statex // --- Root Node: Selector (?) ---
\Procedure{Robot\_Decision\_Cycle}{$s_i, T, W_i$}
    \If{\Call{Act\_On\_Task}{$s_i, T, W_i$} == SUCCESS} \Comment{1. Branch: Highest priority}
        \State \textbf{return}
    \EndIf
    
    \If{\Call{Acquire\_Task}{$s_i, T, W_i$} == SUCCESS} \Comment{2. Branch: Medium priority}
        \State \textbf{return}
    \EndIf
    
    \State \Call{Execute\_Wander\_Behavior}{} \Comment{3. Branch: Fallback safety behavior}
    \State \textbf{return}
\EndProcedure
\end{algorithmic}
\end{algorithm}

\begin{algorithm}[H]
\caption{BT Helper Functions}
\label{alg:helper_functions}
\begin{algorithmic}[1]
\Statex // --- 1. Branch: Act on Task (Sequence $\rightarrow$) ---
\Function{Act\_On\_Task}{$s_i, T, W_i$}
    \If{robot $i$ has a valid target $t_{current}$ AND is the winner} \Comment{Condition: "Is Target Valid?"}
        \State Execute action for $t_{current}$ (e.g., Move, Inspect, Rescue) \Comment{Action: "Execute Task"}
        \State \textbf{return} SUCCESS \Comment{Or RUNNING, ending the tick}
    \EndIf
    \State \textbf{return} FAILURE \Comment{Sequence fails, try next branch}
\EndFunction

\Statex
\Statex // --- 2. Branch: Acquire Task (Sequence $\rightarrow$) ---
\Function{Acquire\_Task}{$s_i, T, W_i$}
    \State $B_i \leftarrow \text{Build\_Bundle}(s_i, T, W_i)$ \Comment{Action: "Build Bundle"}
    \If{$B_i$ is empty}
        \State \textbf{return} FAILURE
    \EndIf
    
    \State $\text{Run\_Conditional\_Consensus}(s_i, B_i, W_i)$ \Comment{Action: "Run Consensus"}
    \State $t_{new} \leftarrow \text{Select\_New\_Target}(B_i, W_i)$ \Comment{Action: "Select Target"}
    
    \If{$t_{new}$ is valid}
        \State Set $t_{new}$ as current target
        \State \textbf{return} SUCCESS \Comment{New target acquired, tick ends}
    \EndIf
    \State \textbf{return} FAILURE \Comment{Sequence fails, try next branch}
\EndFunction

\Statex
\Statex // --- 3. Branch: Fallback (Action) ---
\Procedure{Execute\_Wander\_Behavior}{}
    \State Execute Wander behavior
    \State \textbf{return} SUCCESS \Comment{Fallback always succeeds}
\EndProcedure
\end{algorithmic}
\end{algorithm}

As formalized in Algorithm \ref{alg:main_cycle_bt} and Algorithm \ref{alg:helper_functions}, the agent's decision cycle encapsulates the prioritized logic of the Behavior Tree. The main \texttt{Robot\_Decision\_Cycle} (Algorithm \ref{alg:main_cycle_bt}) acts as the root selector, attempting to execute branches in order of priority. The cycle begins by calling the highest-priority branch, \texttt{Act\_On\_Task} (Algorithm \ref{alg:main_cycle_bt}, line 4). This function checks for a pre-existing and valid task assignment (Algorithm \ref{alg:helper_functions}, line 2), ensuring task persistence and preventing the agent from abandoning its objective unnecessarily.

If this branch returns \texttt{FAILURE} (e.g., no valid task), the root selector transitions to the proactive task acquisition branch, \texttt{Acquire\_Task} (Algorithm \ref{alg:main_cycle_bt}, line 7). This function, detailed in Algorithm \ref{alg:helper_functions}, invokes the \texttt{Build\_Bundle} procedure (line 9) for local task evaluation and the \texttt{Run\_Conditional\_Consensus} procedure (line 13) for efficient, event-triggered team coordination. If this process results in a successful new assignment, the agent commits to the new target and the cycle ends.

Should both higher-priority branches fail, the root selector executes the final fallback behavior, \texttt{Execute\_Wander\_Behavior} (Algorithm \ref{alg:main_cycle_bt}, line 10). This final step acts as a crucial safety net, ensuring the robot is never idle and can continue to contribute to the mission's exploratory goals, thereby guaranteeing system robustness.

\subsection{Local Plan Formation (Bundle Construction)}
The first step in the task acquisition process is the formation of a local plan, known as a "bundle." This phase is executed independently by each robotic agent, relying solely on its own state and sensory perception of the environment. The objective is for each robot to autonomously identify available tasks and construct a prioritized, ordered sequence of these tasks it intends to pursue. This bundle represents the agent's local, greedy plan, which forms the basis for the subsequent negotiation and conflict resolution during the consensus phase.

The core of this process is the calculation of a utility score, $U_{ij}$, which quantifies the value or reward for robot $i$ successfully completing task $j$. The utility function is designed to incorporate the most critical factors for efficient decision-making in our scenario. The primary component of the score is proximity, as assigning tasks to the nearest available agents minimizes travel time and energy consumption. This is modeled as the inverse of the distance $d_{ij}$ between the robot and the task. The second, and equally crucial, component is the agent's suitability for the task, which encapsulates the system's heterogeneity. In our scenario, this is determined by color compatibility (we assume that this information is available for the consensus-based strategies with a "bid-then-verify" challenge, where agents must still execute a formal inspection to confirm the assignment). If a robot's capability (color) does not match a known task's requirement, the utility score is multiplied by a significant penalty factor of 0.1 to deprioritize inefficient assignments. The complete utility function is therefore defined as:
\begin{equation}
U_{i}(j) =
\begin{cases}
    \frac{1}{d_{ij} + \epsilon} & \text{if colors match} \\
    0.1 \times \frac{1}{d_{ij} + \epsilon} & \text{if colors do not match}
\end{cases}
\label{eq:utility}
\end{equation}
where $\epsilon$ is a small constant to prevent division by zero.

With a method to score every potential task, the robot greedily assembles its bundle by iteratively selecting the available task with the highest utility score. The size of this bundle is typically limited by a parameter $L_i$, which restricts the number of tasks an agent can plan for at any given time. This limitation is critical for maintaining the agent's focus on the most immediate, high-value objectives and for ensuring that the subsequent consensus process remains computationally tractable, as it reduces the amount of information that needs to be communicated and processed. The entire greedy selection process for constructing the task bundle is formally detailed in Algorithm 3.

\begin{algorithm}[H]
\caption{Build\_Bundle}
\label{alg:build_bundle}
\begin{algorithmic}[1]
\State \textbf{Input:} Robot state $s_i$, Task list $T$, World knowledge $W_i$
\State \textbf{Output:} New task bundle $B_i$

\State Initialize $B_i \leftarrow \emptyset$
\State $T_{available} \leftarrow T \setminus \{\text{tasks already in a bundle}\}$
\While{$|B_i| < L_i$ \textbf{and} $T_{available} \neq \emptyset$}
    \State $j^* \leftarrow \text{null}$, $U_{max} \leftarrow -\infty$
    \For{each task $j \in T_{available}$}
        \State Calculate utility $U_{ij}$ for task $j$ using Equation \ref{eq:utility}
        \If{$U_{ij} > U_{max}$}
            \State $U_{max} \leftarrow U_{ij}$
            \State $j^* \leftarrow j$
        \EndIf
    \EndFor
    \If{$j^* \neq \text{null}$}
        \State Add task $j^*$ and its bid $U_{max}$ to $B_i$
        \State $T_{available} \leftarrow T_{available} \setminus \{j^*\}$
    \Else
        \State \textbf{break}
    \EndIf
\EndWhile
\State \textbf{return} $B_i$
\end{algorithmic}
\end{algorithm}

This algorithm ensures that each agent generates a locally optimal plan based on its current perspective and capabilities. This bundle forms the basis of the agent's intentions, which will subsequently be communicated and deconflicted during the event-triggered consensus phase, described in the following section.

\subsection{Intelligent Consensus by Events (Event-Triggered Consensus)}
A core innovation of our CBBA-ETC framework is its departure from traditional, periodic communication protocols. Instead, it employs an Event-Triggered Control (ETC) mechanism that serves as the cornerstone of its adaptive communication strategy. This approach intelligently manages the inherent trade-off between the quality of the task-assignment solution and the communication overhead required to achieve it. The fundamental principle is that consensus is not a routine, scheduled process but a strategic action, initiated only when significant new information arises that has a high probability of beneficially altering the collective task allocation. This prevents the network saturation and inefficient resource consumption characteristic of more naive reactive or periodic approaches, which is a critical capability in resource-constrained missions.

The decision to trigger a consensus round is governed by a set of local, event-based heuristics evaluated within each agent's Behavior Tree. These heuristics represent the occurrence of strategically relevant events that justify the cost of communication. The system evaluates four primary trigger conditions:

\begin{itemize}
    \item \textbf{$Tri_{init}$ (Initial Plan Formation):} If an agent formulates a new, non-empty task bundle, a consensus round is triggered to announce these initial intentions to the swarm. This ensures new plans are immediately shared for deconfliction.
    \item \textbf{$Tri_{\Delta bid}$ (Significant Bid Change):} If an agent's own utility score (bid) for its primary task changes by more than a predefined threshold, $\theta_{bid\_change}$, it initiates consensus. This allows the agent to react to significant changes in its own state or its perception of the task's value.
    \item \textbf{$Tri_{conflict}$ (High-Value Conflict Opportunity):} An agent triggers consensus if it identifies an opportunity to outbid a known winner for a high-value task by a significant margin, $\theta_{outbid\_margin}$. This enables proactive resolution of high-potential conflicts.
    \item \textbf{$Tri_{fallback}$ (Adaptive Time-Based Synchronization):} As a critical safeguard, consensus is initiated if no other event has occurred for a duration exceeding a dynamically adapting time interval, $I_{adapt}$. This mechanism prevents the swarm from falling out of sync during periods of low event activity and guarantees that the consensus error remains bounded.
\end{itemize}

Once any of these conditions trigger the event, the agent engages in the consensus process. Robots communicate their intentions (bundles and associated bids) to their neighbors. Through this local information exchange, conflicts are resolved, with the standard CBBA protocol dictating that the robot with the highest bid wins the contested task. A deterministic tie-breaking rule (e.g., lowest robot ID) ensures unique assignments. Each agent then updates its local view of the global assignments. This entire process is configured to be completed in a single communication round to facilitate rapid decision-making. The logic for evaluating the trigger conditions and executing the consensus protocol is formalized in Algorithm 4.

\begin{algorithm}[H]
\caption{Run\_Conditional\_Consensus }
\label{alg:cond_consensus}
\begin{algorithmic}[1]
\State \textbf{Input:} Robot state $s_i$, Bundle $B_i$, World knowledge $W_i$ (includes winning bids $y$, winners $z$, adaptive interval $I_{adapt}$, last consensus time $t_{last}$)

\Statex // --- Evaluate Event-Trigger Conditions ---
\State $trigger_{init} \leftarrow \text{CheckNewPlan}(B_i)$ \Comment{$Tri_{init}$ }
\State $trigger_{\Delta bid} \leftarrow \text{CheckBidChange}(s_i, B_i, \theta_{bid\_change})$ \Comment{$Tri_{\Delta bid}$ }
\State $trigger_{conflict} \leftarrow \text{CheckConflictOpportunity}(s_i, W_i, \theta_{outbid\_margin})$ \Comment{$Tri_{conflict}$ }
\State $trigger_{fallback} \leftarrow \text{CheckFallbackTimer}(t_{last}, I_{adapt})$ \Comment{$Tri_{fallback}$ }

\Statex
\If{$trigger_{init} \lor trigger_{\Delta bid} \lor trigger_{conflict} \lor trigger_{fallback}$} \Comment{Run if any trigger is true }
    \Statex // --- Perform Consensus ---
    \State Broadcast local bundle $B_i$ and winning bids $y_i$ to neighbors 
    \State Receive bundles and bids from neighbors
    \State Update local knowledge $(y_i, z_i)$ based on highest bids 
    \State Resolve conflicts for tasks in $B_i$ (remove if outbid) 
    \State $t_{last} \leftarrow \text{Time.now}$
    
    \Statex // --- Adapt Consensus Interval ---
    \State $L_c \leftarrow \text{number of tasks lost in this consensus round}$ \Comment{Proxy for conflict level }
    \State Calculate $\Delta L_c$ from previous round 
    \State $I_{adapt} \leftarrow \text{AdaptConsensusInterval}(L_c, \Delta L_c, I_{adapt})$ \Comment{Update rule }
\EndIf

\Statex // --- Prepare for next cycle ---
\State Update $U^{prev}$ with current bids from $B_i$ \Comment{Store bids for next $Tri_{\Delta bid}$ check }
\end{algorithmic}
\end{algorithm}

Algorithm \ref{alg:cond_consensus} encapsulates the core of the system's communication intelligence. The process begins by evaluating the four event-trigger conditions (lines 2-5), which are abstracted into helper functions. These triggers represent the agent's local heuristics for deciding when to communicate:
\begin{itemize}
    \item \textbf{$Tri_{init}$ (\texttt{CheckNewPlan})}: Triggers if the agent has just created a new, non-empty task bundle $(B_i \text{ is new and not empty})$.
    \item \textbf{$Tri_{\Delta bid}$ (\texttt{CheckBidChange})}: Triggers if the agent's own utility score for its primary task ($j_p$) changes significantly compared to its previously recorded score $ (|U_i(j_p) - U_i^{prev}(j_p)| > \theta_{bid\_change})$.
    \item \textbf{$Tri_{conflict}$ (\texttt{CheckConflictOpportunity})}: Triggers if the agent identifies a high-value opportunity, specifically if its bid for a potential task ($j_k$) is high enough to outbid the current known winner $y_k(j_k)$ by a specified margin $(U_i(j_k) > y_k(j_k) + \theta_{outbid\_margin})$.
    \item \textbf{$Tri_{fallback}$ (\texttt{CheckFallbackTimer})}: Acts as a safeguard, triggering if the time elapsed since the last consensus ($t_{last}$) exceeds the dynamically adapting interval $I_{adapt}$ $(\text{Time.now} - t_{last} \ge I_{adapt})$.
\end{itemize}
If any of these conditions are met (line 6), indicating a strategically valuable moment for coordination, the agent proceeds to engage in the distributed consensus protocol (lines 7-11). A crucial element of the framework's adaptability is nested within this block: following a consensus round, the agent immediately re-evaluates and adjusts its adaptive fallback interval, $I_{adapt}$, based on the level of conflict it just experienced ($L_c$) (lines 12-14). This creates an intelligent feedback loop, allowing the swarm to self-regulate its coordination pace. Finally, the agent records its current bids (line 16) to enable the stateful evaluation of triggers like $Tri_{\Delta bid}$ in the subsequent decision cycle.

\subsection{Adaptive Consensus Interval Mechanism}
A critical component of the event-triggered framework's intelligence and efficiency is the mechanism that governs the fallback trigger, $Tri_{fallback}$. Rather than relying on a static, predefined time interval, the CBBA-ETC system employs a dynamic interval, $I_{adapt}$, that is continuously adjusted based on the perceived level of conflict within the multi-robot team. This allows the swarm to achieve a form of collective self-regulation, intelligently modulating its coordination frequency to match the stability of the operational environment. During periods of high contention and dynamic change, the swarm automatically increases its coordination pace (by shortening $I_{adapt}$) to resolve conflicts quickly. Conversely, during stable periods with low conflict, it reduces the frequency of fallback consensus rounds (by lengthening $I_{adapt}$) to conserve communication bandwidth and energy resources.

This adaptation is managed by the `AdaptConsensusInterval` component within the robot's Behavior Tree, which learns from the outcomes of recent consensus rounds. After each consensus round, the agent evaluates the number of tasks it lost, $L_c(t)$, which serves as a direct proxy for the level of conflict it experienced. This value, along with its change from the previous round, $\Delta L_c(t)$, is used to adjust $I_{adapt}$. The adjustment logic is governed by a continuous, trend-aware function that provides smooth adaptation and integrates safety bounds explicitly. The update rule is expressed as:
\begin{equation}
I_{adapt}(t) = \text{clip}\left(I_{adapt}(t-1) \times \rho_{adaptive}(L_c(t), \Delta L_c(t)), I_{base} \times \phi_{min}, I_{base} \times \phi_{max}\right)
\end{equation}
Here, $\rho_{adaptive}$ is a dynamic adjustment factor that incorporates both the current conflict level and its trend. It is defined as:
\begin{equation}
\rho_{adaptive}(L_c, \Delta L_c) = \rho_{base} - \omega(L_c, \Delta L_c)
\end{equation}
The adjustment is driven by the function $\omega(L_c, \Delta L_c)$, which is defined piecewise based on high and low conflict thresholds, $\theta_{high}$ and $\theta_{low}$:
\begin{equation}
\omega(L_c, \Delta L_c) = 
\begin{cases}
    \kappa \cdot \sigma(L_c - \theta_{high}) \cdot (1 + \gamma \cdot \text{sign}(\Delta L_c)) & \text{if } L_c \ge \theta_{high} \\
    0 & \text{if } \theta_{low} < L_c < \theta_{high} \\
    -\lambda \cdot \sigma(\theta_{low} - L_c) \cdot (1 + \gamma \cdot \text{sign}(-\Delta L_c)) & \text{if } L_c \le \theta_{low}
\end{cases}
\end{equation}
where $\sigma(x)$ is the sigmoid function, which ensures smooth transitions and avoids abrupt behavioral changes.

The design of this formulation reveals a deliberate philosophy aimed at creating a stable yet highly adaptive control system. A key feature is the calibrated asymmetry between the parameters $\kappa$ and $\lambda$. The system reacts conservatively to an \textit{increase} in conflict (a low $\kappa$ value slowly decreases the interval to avoid message storms), but it acts aggressively to capitalize on periods of \textit{stability} (a high $\lambda$ value rapidly increases the interval to maximize resource savings). Furthermore, the "dead zone" between the thresholds provides hysteresis, ensuring the system remains robust by not overreacting to minor conflict fluctuations. The inclusion of the conflict trend, $\Delta L_c$, allows for a more nuanced and proactive adaptation, while the explicit `clip()` function enforces safety bounds to guarantee that the coordination frequency always remains within an acceptable operational range.

This adaptive mechanism is a crucial enabler of the swarm's emergent intelligence, empowering the collective to autonomously balance mission responsiveness with resource conservation.

\subsection{Behavior Trees (BTs) for Robust Task Execution and Contingency Management}

Behavior Trees are central in CBBA-ETC for translating the abstract task assignments derived from the CBBA consensus process into concrete robot actions and for managing the complexities of task execution at the individual agent level. BTs provide a structured, hierarchical, and reactive control flow \cite{ogrenBehaviorTreesRobot2022}. In CBBA-ETC, the BT defines a robot's operational cycle conceptually as follows:

\begin{enumerate}
\item \textbf{Target Validation and Action:}~If a task is currently assigned via CBBA, the BT first validates its continued relevance and the robot's assignment status. If valid, it executes a sequence for task completion, which may involve navigation to the target, inspection to gather further information (e.g., tasks's color), and performing the specific rescue or abandonment action based on this information.
\item \textbf{Task Acquisition (CBBA Process Invocation):}~If no valid target is currently assigned, or if an assignment becomes invalidated (e.g., due to losing the task in a consensus round), the BT logic directs the robot to initiate the CBBA process (bundle building, conditional consensus, and new target selection).
\item \textbf{Exploratory Behavior:}~As a fallback, if no task is assigned and the CBBA process does not yield a new assignment, the BT triggers an exploratory behavior (e.g., wandering) to search for new tasks or information.
\end{enumerate}

\begin{figure}
  \begin{center}
    \includegraphics[width=0.95\textwidth]{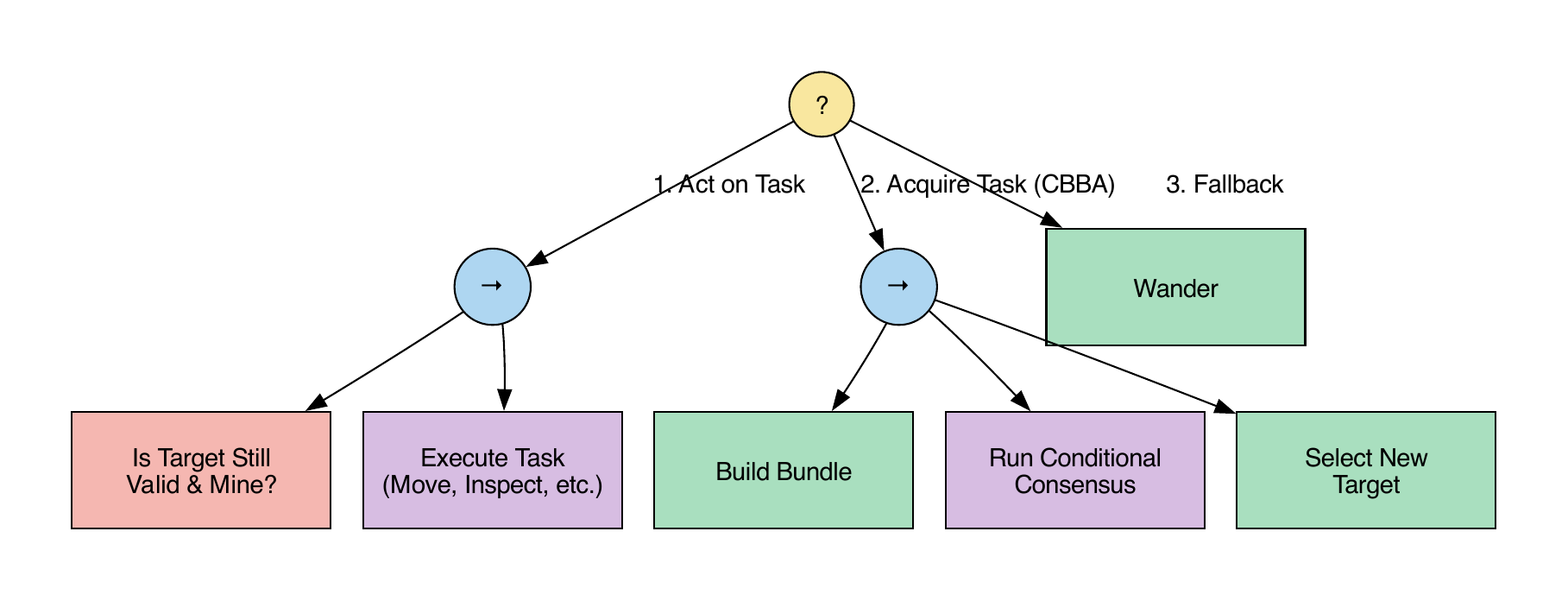}
  \end{center}
  \caption{This figure illustrates the robot's high-level decision-making architecture, implemented as a Behavior Tree (BT). The tree processes logic from the root downwards and from left to right, enabling a prioritized system of behaviors. Composition Nodes are represented by circles, Leaf Nodes by rectangles, conditions by red, and actions by green or purple (grouped actions)}\label{fig:btbase}
\end{figure} 

The presented diagram of figure \ref{fig:btbase} illustrates the robot's high-level decision architecture, implemented as a Behavior Tree (BT). The logic of a BT is processed from the root node downwards and from left to right, which naturally allows for the creation of a prioritized system of behaviors \cite{ogrenBehaviorTreesRobot2022}. The nodes in the diagram can be classified into two main types:

\begin{itemize}
\item Composition Nodes (Circles). These are the internal nodes that direct the flow of execution. They don't perform actions themselves but orchestrate their child nodes. There are two types in this tree:
  \begin{itemize}
  \item \textbf{Selector (?)}: Also known as a ``Fallback'' or ``Priority'' node. It attempts to execute its children in order (from left to right) until one of them succeeds. The root node is a Selector, meaning the robot will always try to ``Act on Task'' first before attempting to ``Acquire New Task'', and so on.
  \item \textbf{Sequence (→)}: Executes its children in order, one after another. It only succeeds if all its children succeed. If one of the children fails, the entire sequence immediately fails. It's ideal for defining step-by-step processes, such as the ``Act'' and ``Acquire'' branches.
  \end{itemize}
\item Leaf Nodes (Rectangles) These are the nodes that perform the actual work. They have no children and return a success or failure state. In the diagram, they represent:
  \begin{itemize}
  \item \textbf{Conditions (red)}: They check a state of the robot or the world, such as ~\textbf{Is Target Still Valid \& Mine?}
  \item \textbf{Actions (green/purple)}: They execute a specific task, such as~\textbf{Wander}~or~\textbf{Build Bundle}. Purple nodes represent sub-trees or more complex actions that have been grouped to simplify the view.
  \end{itemize}
\end{itemize}

The robot's logical flow is therefore clear and robust: the root~\textbf{Selector (?)}~first attempts to execute the~\textbf{Sequence (→)} ''Act on Task.'' If this fails (for example, because the~\textbf{Is Target Still Valid?}~condition is not met), the selector moves on to the second branch, the~\textbf{Sequence (→)}~``Acquire Task''. If this also fails (for example, because the CBBA process doesn't find any new tasks), the selector finally executes the last option, the~\textbf{Action}~``Wander'', which acts as the robot's default behavior.

This structured, prioritized execution model allows for both deliberate action towards assigned goals and reactive adaptation when goals change or new opportunities arise. A key advantage of using BTs is the ability to manage local contingencies during task execution. The inherent conditional logic and execution flow of BTs enable robust responses to common operational issues:

\begin{itemize}
\item \textbf{Action Execution Failures:}~Robot actions such as movement, inspection, or rescue may fail with certain probabilities (\(P_{move\_fail}, P_{inspect\_fail}, P_{rescue\_fail}\) respectively). The BT's structure (e.g., sequences, selectors, decorators like ``retry until success'' or ``fallback'') determines how the robot responds to such failures, whether by re-attempting the action, trying an alternative, or abandoning the current sub-task and potentially re-evaluating the overall task.
\item \textbf{Dynamic Target Status:}~Condition nodes within the BT continuously monitor the status of the assigned target (e.g., Has it been completed by another robot? Is it still valid?). If a target's status changes significantly, the BT can interrupt the current action sequence and trigger a re-evaluation, possibly leading back to the CBBA process to seek a new assignment.
\end{itemize}

This capacity for local adaptation and failure recovery within the BT framework makes individual agents more resilient and the overall system more robust to the uncertainties of the environment.

\begin{figure}
  \begin{center}
    \includegraphics[width=0.95\textwidth]{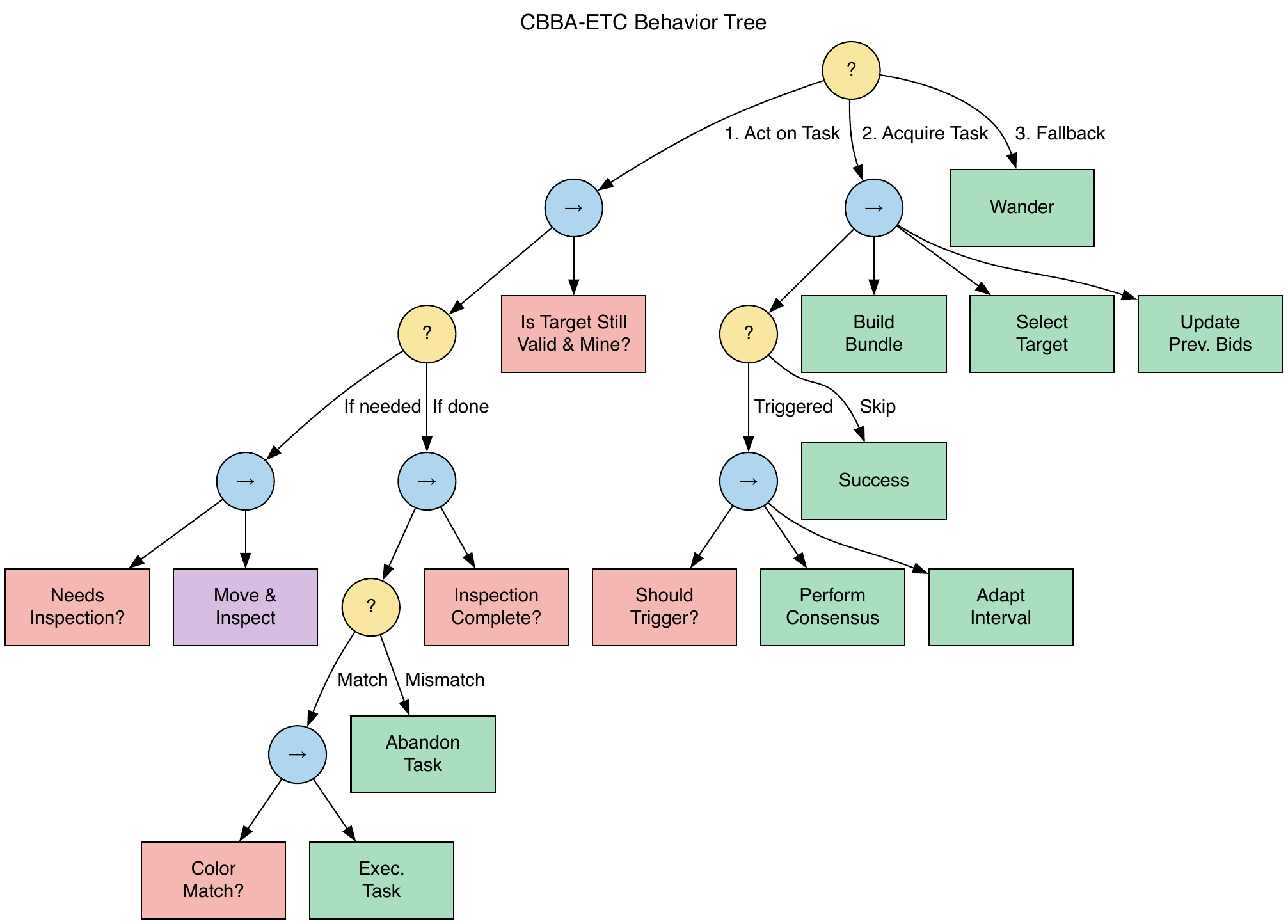}
  \end{center}
  \caption{This figure presents a detailed view of the robot's Behavior Tree (BT) architecture. It reveals the sophisticated contingency logic and resource control that enable local adaptation and failure recovery, making the overall system robust. Some previously simplified actions are expanded to show their complex sub-structures and the underlying intelligence of the robot's behaviors. Composition Nodes are represented by circles, Leaf Nodes by rectangles, conditions by red, and actions by green or purple (grouped actions)}\label{fig:btexpand}
\end{figure}

Following a more detailed diagram is presented in figure \ref{fig:btexpand}. It offers a much more faithful view of the actual code implementation, revealing the sophisticated contingency logic and resource control that BTs bring to the system. In this expanded tree, nodes previously shown as simple actions are broken down into their own sub-structures, exposing the intelligence of the robot's behavior. 

The ``\textbf{Act on Task}'' branch, for instance, isn't just a two-step sequence. It's revealed to be controlled by a sub-selector that first decides whether the task needs to be inspected or if it can be acted upon immediately. If inspection is needed, a sequence is executed that includes moving toward the target and performing the inspection action. If inspection is already complete, another sub-selector decides between two final sequences: one for~\textbf{rescuing}~(if the~\textbf{Color Match?}~condition succeeds) and another for~\textbf{abandoning}~(if the color condition fails). This nested structure of selectors and sequences demonstrates how the robot can manage a complex workflow with multiple decision and contingency points in a modular and clear way.

Similarly, the subgraph detailing the ``\textbf{CBBA-ETC Process}'' shows that task acquisition is more than a simple sequence of three actions. The central element is a~\textbf{Decision Selector}~that implements Event-Triggered Control (ETC) logic. This selector chooses between executing the full consensus sequence or skipping it if no trigger condition (\textbf{Should Trigger?}) has been met. The consensus sequence itself is broken down into its three key components: checking the trigger condition, executing the consensus algorithm, and adapting the time interval for the next cycle. This structure explicitly visualizes how the rules for adaptive communication are directly integrated into the robot's behavior flow, enabling efficient and intelligent coordination.

\subsection{Architectural Principles for Efficient, Event-Driven Communication}

The high communication efficiency of the CBBA-ETC protocol is not the result of a single component, but emerges from a set of interconnected architectural principles. These principles govern how individual agents act, adapt, and utilize the communication network based on local, high-value information. The interplay between these mechanisms is the foundation for the system's collective ability to perform efficient, resilient, and adaptive task allocation while minimizing network load.

\begin{itemize}
    \item \textbf{Selective Network Utilization via Event-Triggered Communication:} By initiating consensus only for high-value events—such as a significant change in task utility ($Tri_{\Delta bid}$) or a direct conflict over a high-priority task ($Tri_{conflict}$)—the system collectively avoids network saturation and conserves critical energy resources. This represents a fundamental principle of resource-aware coordination, allowing the distributed system to focus its limited bandwidth on information that is most likely to improve the collective strategy.

    \item \textbf{System-Level Self-Regulation of Communication Pace:} The adaptive fallback interval ($I_{adapt}$) enables system-level self-regulation of the network's coordination frequency. The ability to dynamically adjust this pace based on observed environmental conflict ($L_{c}$) allows the system to maintain both stability and efficiency across diverse conditions. It automatically increases its coordination frequency during periods of high conflict and conserves network resources during stable phases, a key feature for adaptive distributed systems.

    \item \textbf{Information Valuation as a Precondition for Communication:} The utility function ($U_{i}(j)$) serves as a mechanism for information valuation, which is the basis for efficient distributed decision-making. By encoding task compatibility and proximity into a local score, each agent can assess whether its local information has sufficient value to warrant a network broadcast and potentially trigger a consensus round. When these high-value local assessments are shared through the consensus process, they lead to an emergent, globally efficient task allocation that respects agent heterogeneity while minimizing low-value communication.
    
    \item \textbf{Network Stability through Local Fault Tolerance:} The Behavior Tree framework provides the foundation for local fault tolerance, which is critical for overall network stability and efficiency. By structuring task execution and providing built-in mechanisms for local contingency management (e.g., retrying a failed action), agents can handle common operational failures autonomously without triggering a system-wide re-coordination event. This crucial separation of local tactical error handling from global strategic planning prevents minor individual setbacks from generating unnecessary network traffic and causing systemic instability.

\end{itemize}

The interplay between these mechanisms is mutually reinforcing: robust local fault tolerance reduces the number of superfluous event triggers, which in turn allows the system's self-regulating communication protocol to remain highly efficient. In this way, an effective and adaptive network coordination strategy emerges from the application of these interconnected, local principles.

\section{Experimental Setup and Evaluation\label{sections:Experimental}}

This section outlines the experimental methodology employed to evaluate the proposed CBBA-ETC algorithm against several baseline multi-robot coordination strategies. We detail the simulation environment, the specifics of the compared algorithms, the experimental scenarios designed to test performance and robustness, the metrics used for evaluation, and the statistical methods applied for result analysis.

\subsection{Simulation Environment Description}

As established, effective swarm behavior hinges on solving the Multi-Robot Task Allocation (MRTA) problem, particularly in dynamic scenarios such as logistics or disaster response, where tasks emerge unexpectedly and robot failures are common. To rigorously evaluate the proposed coordination strategies under these exact conditions, we have developed a custom-built, simulated Search and Rescue (SAR) environment. This testbed was not chosen arbitrarily; it was specifically designed to be a challenging domain that embodies the core algorithmic complexities of MRTA, \textbf{focusing on task allocation under uncertainty, team heterogeneity and time-criticality}.

The environment is a deliberate abstraction intended to isolate these coordination challenges. For instance, the absence of physical obstacles is based on the operational model of Unmanned Aerial Vehicles (UAVs) in open airspace, where pathfinding is often trivial. This design choice shifts the primary challenge from navigation to the core MRTA problem: deciding which agent should pursue which task. 

The simulation environment is a custom-built Search and Rescue (SAR) testbed designed to model task allocation under uncertainty, team heterogeneity, and time-criticality. The environment is a circular arena where a team of functionally heterogeneous robotic agents ($N_R$ robots) operate. The tasks are specifically victims that appear dynamically at unpredictable locations and are characterized by a specific color requirement ($C_{wall}$). The core challenge is the dynamic assignment of tasks to robots with matching capabilities (colors) to maximize the total number of successfully completed tasks within a finite mission duration. Tasks are time-sensitive and will expire if not completed within a finite lifetime ($V_{LIFETIME}$), introducing temporal urgency. Furthermore, the capability requirement of a task is not known in advance and must be discovered by a robot performing a dedicated "inspection" action at close range.

More specifically, the key features of our SAR simulation that model these MRTA complexities are detailed below, followed by the specific parameters used in the baseline experimental scenario.

\begin{itemize}
    \item \textbf{Arena and Agents:} The environment is a circular arena with a radius of \(R_{\text{arena}}=32\,\text{m}\). A team of \(N_{R}\) robots operates within this area, moving at a maximum speed of \(40\,\text{km/h}\).

    \item \textbf{Victim Dynamics:} $N_V$ Victims (tasks) appear at random locations and are characterized by a \emph{wall color} from \(C_{\text{wall}} \in \{\text{RED, GREEN, BLUE}\}\). The simulation supports two distinct operational modes: a steady-state mode where victims are \emph{replaced} upon rescue or expiration (ensuring a continuous task flow), and a finite-task mode where an initial set of victims is \emph{not replaced}. In dynamic scenarios, victims have a finite lifetime of \(V_{\text{LIFETIME}}=100\,\text{s}\), introducing temporal urgency.

    \item \textbf{Robot Heterogeneity and Actions:} Agent heterogeneity is enforced by a color-matching rule: a robot with color $C_{\text{robot}_r}$ can only rescue a victim with a matching wall color. This requires a close-range (\(D_{\text{detection}}=5\,\text{m}\)) ``inspect'' action to discover the wall color. If the colors match, the robot can perform a ``rescue'' action. If they do not match, the robot executes an ``abandon'' action and internally records the victim as incompatible for a period, preventing immediate re-inspection loops.

\item \textbf{Robot Heterogeneity and Actions:} Agent heterogeneity is enforced by a color-matching rule: a robot with color $C_{\text{robot}_r}$ can only rescue a victim with a matching wall color. This requires a close-range (\(D_{\text{detection}}=5\,\text{m}\)) ``inspect'' action to discover the wall color. While this inspection is mandatory for all algorithms, a "bid-then-verify" model is assumed for consensus-based strategies (CBBA, C-CBBA, CBBA-Tree, CBBA-ETC). For these strategies, color information is considered available during the bidding phase to calculate utility, although the robot must still execute the formal ``inspect'' action upon arrival to confirm the assignment. If the colors match, the robot can perform a "rescue" action. If they do not match, the robot executes an ``abandon'' action and internally records the victim as incompatible for a period, preventing immediate re-inspection loops.

    \item \textbf{Robot Perception:} Agents are equipped with a conical sensor defined by a vision range of \(R_{\text{vision}}=12\,\text{m}\) and a vision angle of \(A_{\text{vision}}=\pi/2\,\text{rad}\). Within this field of view, they can detect the presence and location of victims and other robots, but cannot discern the critical wall color from a distance.

    \item \textbf{Communication:} Agents can communicate within a defined range of \(R_{\text{comm}}=57\,\text{m}\). This range ensures direct communication is possible between most agents in the test area but does not guarantee full, end-to-end connectivity across the entire diameter.

    \item \textbf{Operational Uncertainty:} The simulation incorporates stochasticity through probabilistic failures for movement, inspection, and rescue actions ($P_{\text{move\_fail}}, P_{\text{inspect\_fail}}, P_{\text{rescue\_fail}}$), with failed movements incurring a time penalty (\(T_{\text{move\_penalty}}=0.5\,\text{s}\)).
    
    \item  \textbf{Adapt Consensus Interval}: For the experimentation presented in this paper, the consensus parameters were determined through a rigorous empirical tuning process to optimize system performance. The values used are: $\rho_{base}=1.0$, $\theta_{high}=3$, $\theta_{low}=1$, with sigmoid parameter $\mu=10.0$. The scaling factors are asymmetrically calibrated to $\kappa=0.1$ and $\lambda=6.0$, with a trend influence of $\gamma=0.001$. The safety bounds are set by $\phi_{min}=3.0$ and $\phi_{max}=8.0$. 
\end{itemize}

This simulated domain provides a challenging and highly configurable platform to systematically evaluate and compare diverse multi-robot coordination algorithms. Its design ensures that experimental results are a direct consequence of the coordination strategy's ability to manage distributed information, allocate tasks efficiently, and achieve robust performance in a dynamic, time-constrained, and partially observable environment.

To implement this environment, we developed a custom testbed using \emph{SimPy}\footnote{\url{https://simpy.readthedocs.io/}}, a process-based discrete-event simulation framework in Python. A custom solution was chosen over existing high-level robotics simulators like SPACE \cite{jangSPACEPythonbasedSimulator2024} to transparently implement the specialized mechanics of our SAR scenario and its specific stochastic failure models. This foundational approach provides several key advantages: it offers unparalleled control for precise, low-level metric extraction; it is computationally efficient and inherently suited for large-scale parallelization of Monte Carlo analyses; and it ensures that experimental results are a direct consequence of the coordination strategy itself by avoiding the overhead of a more complex simulator. This guarantees the framework is perfectly tailored to the research questions.

To ensure clarity and facilitate the reproducibility of our results, the specific parameters for the baseline scenario (\texttt{base\_R20\_V100}), from which all variations are derived, are summarized in Table \ref{tab:params}.
\begin{table}[h!]
\centering
\caption{Baseline Experimental Parameters (`base\_R20\_V100`)}
\label{tab:params}
\begin{tabular}{l l p{6cm}}
\toprule
\textbf{Parameter} & \textbf{Value} & \textbf{Description} \\
\midrule
\multicolumn{3}{l}{\textbf{Simulation Control}} \\
Simulation Duration ($D$) & 3000s & Total simulation time for each trial. \\
Repetitions ($T$) & 50 & Number of trials conducted for each experimental configuration. \\
\midrule
\multicolumn{3}{l}{\textbf{Environment}} \\
Arena Radius ($R_{arena}$) & 32m & Radius of the circular operational area. \\
Number of Robots ($N_{R}$) & 20 & Total number of agents in the swarm for the baseline scenario. \\
Number of Tasks (victims) ($N_{V}$) & 100 & Initial and steady-state number of victims (with replacement on). \\
Victim Lifetime ($V_{LIFETIME}$) & 100s & Time window before a victim is considered "lost" if not rescued. \\
\midrule
\multicolumn{3}{l}{\textbf{Robot Capabilities}} \\
Max Speed ($S_{max}$) & 40 km/h & Maximum movement speed of the robotic agents. \\
Move Penalty Time ($T_{move\_penalty}$) & 0.5s & Time penalty incurred by a robot after a failed movement action. \\
Communication Range ($R_{comm}$) & 57m & Maximum distance for direct communication between two robots. \\
Vision Range ($R_{vision}$) & 12m & The maximum distance at which a robot can detect victims or other agents. \\
Vision Angle ($A_{vision}$) & $\pi/2$ rad & The field-of-view angle for the robot's sensor cone. \\
Detection Distance ($D_{detection}$) & 5m & The required proximity to a victim to perform an "inspect" action for non consensus-based strategies. \\
\midrule
\multicolumn{3}{l}{\textbf{Stochastic Model (Baseline)}} \\
Action Failure Probability & 0\% & General probability of failure for actions in the baseline scenario. \\
Agent Failure Probability & 0\% & Probability of a robot becoming permanently disabled in the baseline. \\
Packet Loss Probability & 0\% & Probability of a communication message being lost in the baseline. \\
\bottomrule
\end{tabular}
\end{table}

\begin{figure}
  \begin{center}
    \includegraphics[width=0.95\textwidth]{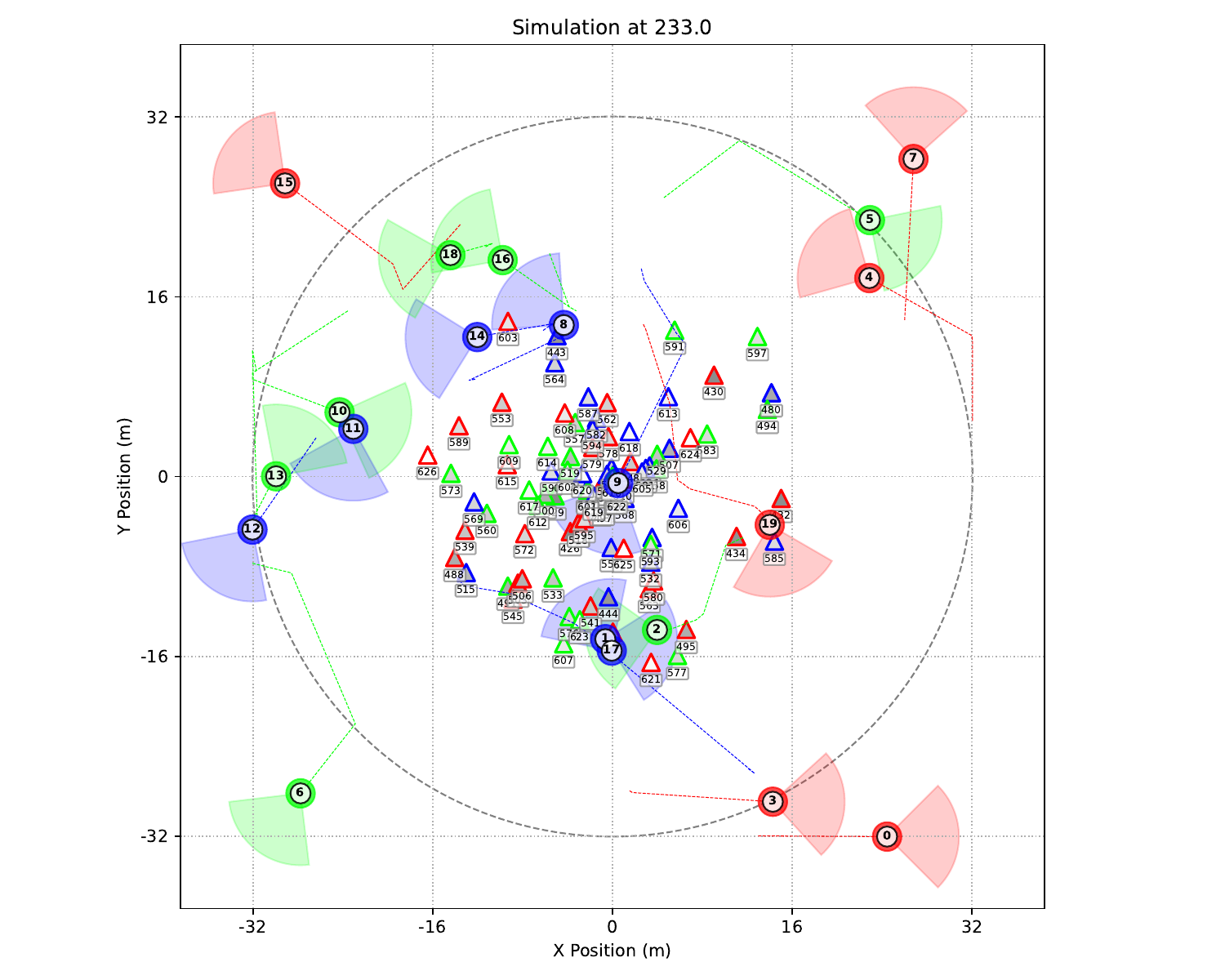}
  \end{center}
  \caption{This figure shows the simulated Search and Rescue (SAR) environment at a specific time step (233 seconds). The circular grey dashed line represents the victims boundary. Robots are depicted as numbered circles with an outer ring indicating their assigned color (red, green, or blue). The colored areas extending from each robot represent their conical sensory perception field. Victims (tasks) are shown as numbered triangles, with their fill color indicating their wall color. Thin dashed lines trace the recent trajectories of the robots. A blue arrow originating from a robot indicates its current movement target.}\label{fig:sim}
\end{figure}

To contextualize the performance of our system in a dynamic and complex scenario, Figure \ref{fig:sim} presents a representative Search and Rescue (SAR) simulation at a specific time step. This visualization details the spatial positions of robots and tasks (victims) within the circular arena, revealing the operational state of the multi-robot team. Key elements depicted include the distribution of robots and victims, the sensory fields of the robotic agents, and the trajectories followed by the robots, providing a qualitative insight into the complexities of real-time task allocation and coordination in this heterogeneous domain.

All the experiments presented in this section are executed for \(D=3000s\) simulation time units, with trials (repetitions) for each configuration equal to 50, to account for stochastic variations.

\subsection{Compared Algorithms}\label{compared-algorithms}

To rigorously evaluate the performance of the proposed~\textbf{CBBA-ETC}~framework, its capabilities are benchmarked against a carefully selected suite of baseline algorithms. This selection is not arbitrary; rather, it represents a graduated progression in coordination and communication complexity, designed to systematically isolate and quantify the benefits of each architectural feature. The comparison begins with the most fundamental baseline,~\textbf{Tree}, a purely reactive and non-communicating agent, to establish a performance floor. From there, the~\textbf{Comm}~algorithm introduces a layer of simple, direct communication to measure the gains from basic information sharing. The analysis then incorporates a formal consensus protocol with the standard~\textbf{CBBA} \cite{han-limchoiConsensusBasedDecentralizedAuctions2009} ~implementation to demonstrate the value of optimized task allocation. Next,~\textbf{CBBA-Tree}~integrates this consensus mechanism into a Behavior Tree framework with a periodic trigger, serving as the direct architectural predecessor to the proposed system. Finally, the approach is compared with Clustering-CBBA (\textbf{C-CBBA}) \cite{dongCommunicationefficientHeterogeneousMultiUAV2025}, a state-of-the-art hierarchical algorithm that systematically groups UAVs to enhance communication efficiency by minimizing the distribution of irrelevant bids and structuring the swarm.

\begin{figure}
  \begin{center}
    \includegraphics[width=0.95\textwidth]{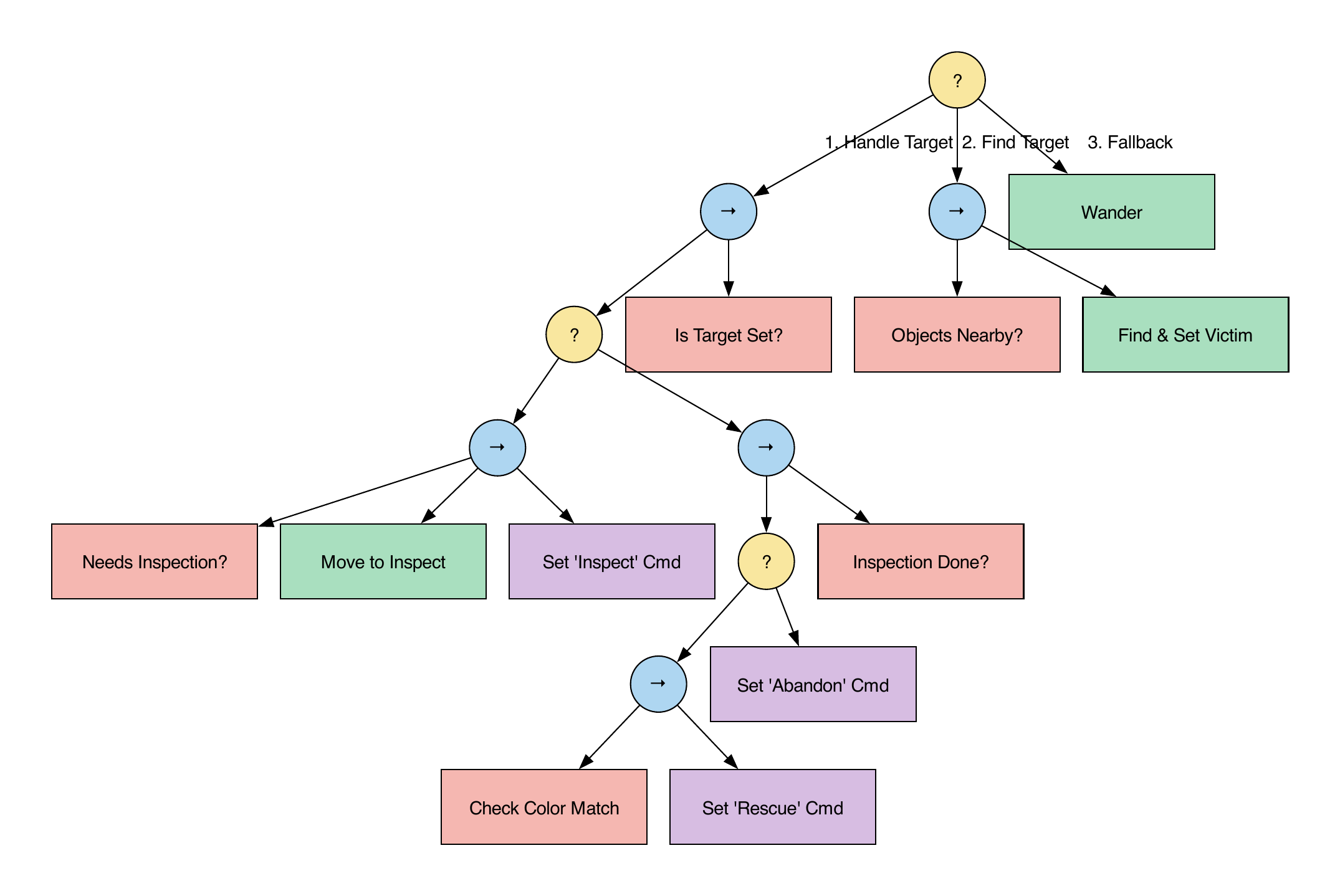}
  \end{center}
  \caption{This figure presents a simple behaviour tree architecture for our SAR problem without communication, called \textbf{Tree}. Composition Nodes are represented by circles, Leaf Nodes by rectangles, conditions by red, and actions by green or purple (grouped actions)}\label{fig:bttree}
\end{figure}

More specifically, the \textbf{Tree} algorithm serves as the fundamental baseline, representing a completely autonomous and non-communicating agent. Each robot's behavior is governed by a Behavior Tree (BT), a hierarchical model that dictates actions based solely on the robot's immediate sensory perception as presented in figure \ref{fig:bttree}. This decision-making structure is strictly prioritized: the robot will first attempt to complete tasks related to its current target. If it has no target, it will then try to identify and select a new one from its local environment. Only if it has no target and cannot find one will it default to a wandering behavior to continue exploring its surroundings. This purely reactive logic ensures the robot is always engaged in a task according to a clear operational hierarchy.

In practice, this model leads to emergent but uncoordinated behavior. When a robot selects a victim, it autonomously proceeds to approach, inspect for compatibility (i.e., the color match), and then commits to either a rescue or abandonment action. When searching for new tasks, it simply chooses the closest available victim without any awareness of the intentions or actions of other robots. The principal limitations of this approach arise directly from this lack of communication. It frequently leads to systemic inefficiencies, such as multiple robots targeting the same victim simultaneously or wasting valuable time inspecting victims that other robots may have already identified as incompatible. Therefore, this algorithm establishes a performance benchmark for an uncoordinated system, against which the benefits of communication and consensus can be clearly measured.

Following, the~\textbf{Comm}~algorithm evolves beyond purely reactive systems by introducing a direct communication layer between robots, while still utilizing a Behavior Tree (BT) architecture. This communication aims to reduce inefficiencies found in uncoordinated models. When a robot targets a victim, it broadcasts its intention to nearby robots. Similarly, if a robot inspects a victim and finds it incompatible (e.g., a color mismatch), it shares this information. This basic cooperative protocol allows robots to announce their targets to avoid redundant efforts and share discoveries to prevent others from making the same inspection errors.

Integrating this information changes the robot's decision-making process. When searching for a new task, the BT now filters potential victims, excluding those already claimed by other robots or known to be incompatible. This leads to smarter target selection and less redundant work. When a comm-type robot selects a victim, it adds that victim to a set of "claimed victims" and communicates this. Other comm robots will then filter these victims, excluding them from consideration as targets. However, this approach has limitations. Communication is reactive and direct, lacking a true consensus mechanism to ensure the most suitable robot is assigned a task; it simply prevents obvious conflicts. Thus, the Comm algorithm demonstrates the clear benefits of simple communication but also highlights the need for more sophisticated task assignment protocols for optimal coordination and efficiency.

The reactive \textbf{CBBA (Consensus-Based Bundle Algorithm)}~baseline introduces a formal, decentralized task allocation mechanism, representing a significant leap in coordination capabilities over the simpler communication model. This algorithm moves beyond simple conflict avoidance to active task negotiation. Instead of merely claiming targets, robots compute a ``bid'' or score for each available victim, quantifying that task's utility (primarily based on proximity, but penalized for known incompatibilities like color mismatches). Each robot first constructs a local ``bundle'' of the best tasks it can perform. Subsequently, during a consensus phase, robots communicate their winning bids to their neighbors, iteratively updating their knowledge until a local agreement is reached, ensuring tasks are allocated to the robots with the highest bids. This algorithm is reactive and includes a periodic fallback. It mainly triggers when its task bundle changes. Nevertheless, if no changes trigger it, it will force a consensus round after 100 seconds.

This process guarantees a more optimal and robust task assignment compared to the simpler ``Comm'' approach. While the Comm model prevents redundant work, CBBA ensures that the most qualified robot (among those in communication range) wins the bid, leading to higher global efficiency. The operational flow is a continuous loop of building a bundle, achieving consensus, and executing the highest-priority task remaining. If a robot loses a bid for a task during consensus, it seamlessly transitions to the next best task in its bundle. This specific implementation is structured as a state machine within the robot's main action cycle rather than a Behavior Tree. Its consensus phase is triggered reactively when a robot's local bundle changes, but it is also~\textbf{periodic}, as a minimum time interval acts as a fallback trigger to ensure the system maintains regular synchronization. In all our implementations of CBBA and its derivatives, if two or more robots submit identical bids for the same task, the conflict will be resolved in favor of the robot possessing the lowest numerical ID. This constitutes a standard deterministic tie-breaking strategy within consensus algorithms.

 The~\textbf{CBBA-Tree}~algorithm represents an architectural fusion, integrating the~\textbf{CBBA (Consensus-Based Bundle Algorithm)}~consensus protocol within the modular framework of a~\textbf{Behavior Tree (BT)}. This hybrid approach serves as the most direct architectural foundation for the proposed CBBA-ETC system. The BT's logic prioritizes action on the currently assigned task, but it introduces a critical control condition: the robot continuously verifies that it remains the legitimate winner of its task according to the latest consensus information. If it loses the task in a negotiation, completes it, or simply doesn't have one assigned, the behavior tree naturally transitions to a lower-priority branch that executes the full CBBA process: building a new bid bundle, running consensus, and selecting a new task.

\begin{figure}
    \begin{center}
      \includegraphics[width=0.95\textwidth]{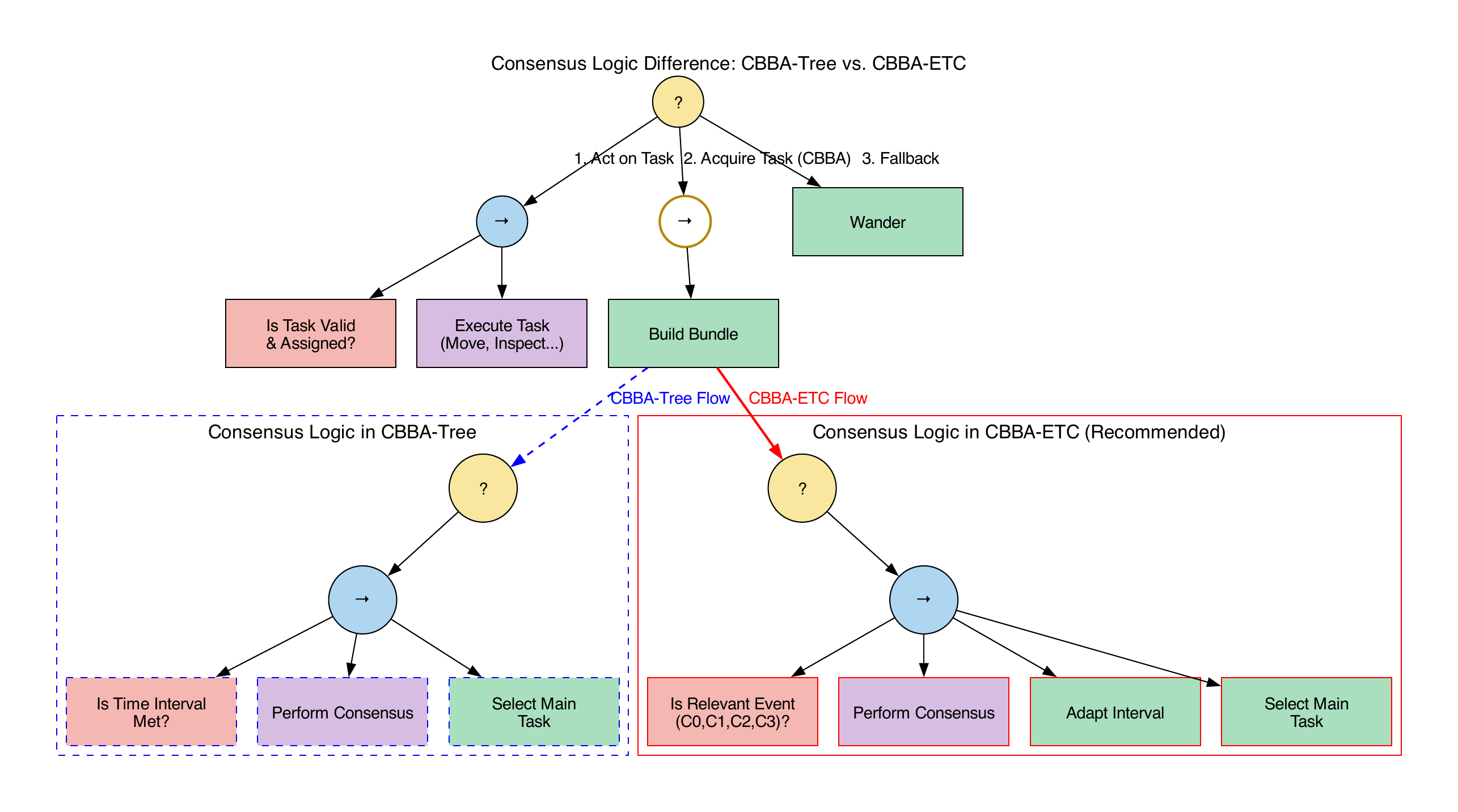}
    \end{center}
    \caption{This figure presents a visual comparison of consensus mechanisms: The diagram highlights how CBBA-Tree (blue/dashed lines) and our CBBA-ETC (red solid) approach task acquisition in a behavior tree, showing periodic vs. event-driven logic}\label{fig:cbbaTreeVsETC}
  \end{figure}

In the~\textbf{CBBA-Tree}~model, consensus is governed by a strictly~\textbf{periodic mechanism}, where in its behavior tree, it checks if at least 100 seconds have passed since the last consensus to initiate a new one. This design choice intentionally simplifies the communication aspect compared to pure CBBA, ensuring that only time dictates when communication occurs. It establishes a clear baseline for comparison by setting communication to happen periodically, with the goal of limiting the amount of information transmitted. As a result, the communication and consensus phase only initiates once a predefined time interval has elapsed, making its communication behavior predictable. However, this predictability comes with a trade-off: the system can be slow to react to urgent environmental changes if they happen just after a consensus round. It may also perform unnecessary communications during periods of low activity. This inefficiency in consensus timing is precisely one of the limitations that the event-triggered mechanism of~\textbf{CBBA-ETC}~aims to resolve. Figure \ref{fig:cbbaTreeVsETC} presents a visual comparison of the  \textbf{CBBA-Tree} consensus compared with the proposed \textbf{CBBA-ETC}.

Finally, \textbf{the Clustering-CBBA (C-CBBA)} algorithm introduces a systematic approach to arranging UAVs into groups according to their preferred tasks in a well-organized manner. This is a state-of-the-art algorithm recognized for communication-efficient task management in heterogeneous clusters. By using an initial bundle structure, communication efficiency is heightened by minimizing the need to distribute irrelevant bids while adhering to the CBBA framework. C-CBBA groups UAVs based on bidding intentions. However, in complex environments with limited communication, where the targets for allocation appear randomly, and communication cannot be guaranteed between UAVs with the same bidding intentions, the grouping strategy may not achieve the expected results.

C-CBBA, is a hierarchical algorithm designed to enhance communication efficiency by structuring the swarm. The core of this approach is to first partition the robots into a predefined number of clusters based on their spatial proximity. For reproducibility, our implementation uses the k-means++ clustering algorithm, consistent with the original paper, with the number of clusters of 2.0, a value determined experimentally in the source research to be optimal. Once clusters are formed, a leader is designated for each one, specifically the robot with the lowest numerical ID within the group. Consensus is then achieved through a two-tiered process. First, an intra-cluster consensus occurs, where non-leader robots unidirectionally transmit their bid information to their respective leaders. This allows each leader to consolidate information and gain situational awareness of its own cluster. Following this, an inter-cluster consensus takes place, where only the leaders communicate bidirectionally among themselves to resolve conflicts at a global level. Once a global consensus is reached, the leaders disseminate the final, conflict-free task assignments back to the robots in their clusters. This hierarchical structure drastically reduces the number of required communication links compared to a flat system where every robot communicates with every other, thereby lowering network overhead.

\subsection{Experimental Scenarios}\label{experimental-scenarios}

The experimental evaluation was designed to rigorously assess the performance of five distinct multi-robot coordination algorithms under a variety of operational conditions, ensuring a comprehensive understanding of their scalability and resilience. The core of the experimentation involves a baseline scenario from which specific parameters are systematically varied. This baseline consists of a 3000s simulation run with 20 robots and 100 tasks (victims), repeated over 50 trials for statistical significance.

A primary focus of the experiments was to evaluate the scalability of each coordination strategy. This was investigated along two main axes: robot density and task density. To assess the impact of swarm size, the number of robots was varied (5, 10, 20, and 40) while the number of tasks was held constant at 100. Conversely, to analyze performance under varying task loads, the number of tasks was adjusted (25, 50, 100, 200, and 500) while the number of robots was fixed at 20. Another scenario altered the fundamental problem structure by disabling victim replacement upon rescue, transforming the simulation from a continuous, steady-state challenge to a finite task-completion problem with 100 initial tasks.

The resilience of the algorithms was tested against several forms of unreliability. To simulate imperfect communication channels, scenarios were introduced with probabilistic packet loss set at 10\% and 30\%, which also included bandwidth limitations to constrain message passing rates. The robustness of the algorithms to hardware unreliability was evaluated by introducing a probability of failure for individual robot actions. These experiments configured the move, inspect, and rescue actions ($P_{move\_fail}, P_{inspect\_fail}, P_{rescue\_fail}$) to fail with equal probabilities of 25\% and 50\%. Finally, system resilience to the complete loss of agents was tested. In these scenarios, individual robots had a chance of permanent failure during the simulation, with probabilities set to 0.01\% and 0.1\% per step after an initial 500s grace period, allowing for an assessment of how well the distributed systems adapt to a dynamically reduced team.

\subsection{Performance Metrics and Statistical Analysis}\label{performance-metrics-and-statistical-analysis}

To evaluate and compare the various coordination strategies, a set of key performance indicators was systematically collected at the conclusion of each simulation trial. The primary measure of effectiveness was the total number of successfully rescued victims, which directly reflects the swarm's ability to complete its main objective. This was complemented by the number of victims that expired before a rescue could be performed, a metric tracked by the environment to quantify missed opportunities. The efficiency and robustness of the algorithms were assessed by tracking the number of failed rescue attempts, which are explicitly counted when a robot engages with a victim but cannot perform the rescue due to an incompatible color key, thus representing wasted effort. Finally, the communication overhead, a critical factor for distributed algorithms, was quantified by two metrics: the total number of messages sent by all robots, and for consensus-based strategies, the total number of negotiation rounds initiated.

To rigorously validate the experimental findings, a formal statistical analysis is performed on the aggregated results from the 50 trials for each scenario. The analysis employs a two-stage process to compare the mean performance of the different algorithms. Initially, a one-way Analysis of Variance (ANOVA) is conducted on each performance metric to determine if any statistically significant differences exist among the group of algorithms. If the ANOVA test yields a significant result, a Dunnett's post-hoc test is subsequently performed. This test is specifically chosen to compare each of the other algorithms directly against a single control group (the~\texttt{cbba-etc}~algorithm in this case) to identify which specific strategies perform significantly better or worse than the advanced baseline. To complement these significance tests, Cohen's $d$ is also calculated for the pairwise comparisons against the control. This provides a measure of the effect size, quantifying the magnitude of the performance difference between algorithms, rather than just its statistical probability.

\section{Results and Discussion\label{section:Results}}

This section presents a comprehensive evaluation of the CBBA-ETC framework, structured to validate its core claims of efficiency, effectiveness, and robustness. We first establish a baseline performance comparison in an ideal scenario, analyzing the critical trade-off between mission effectiveness (tasks completed) and communication cost against all baselines, including the state-of-the-art Clustering-CBBA. Following this, we analyze the framework's scalability by systematically varying both task and robot densities. We then rigorously test the system's robustness against a variety of operational failures, including stochastic action execution failures, communication packet loss, and permanent agent loss, to validate the resilience provided by the integrated Behavior Tree architecture. Finally, we analyze performance in specific SAR-centric scenarios, such as a finite-task problem, to evaluate the framework's ability to efficiently manage team heterogeneity.

\subsection{Baseline Performance: The Effectiveness vs. Efficiency Dilemma}\label{baseline-performance-efficiency-and-effectiveness-in-ideal-conditions}

\begin{figure}
  \begin{center}
    \includegraphics[width=0.7\textwidth]{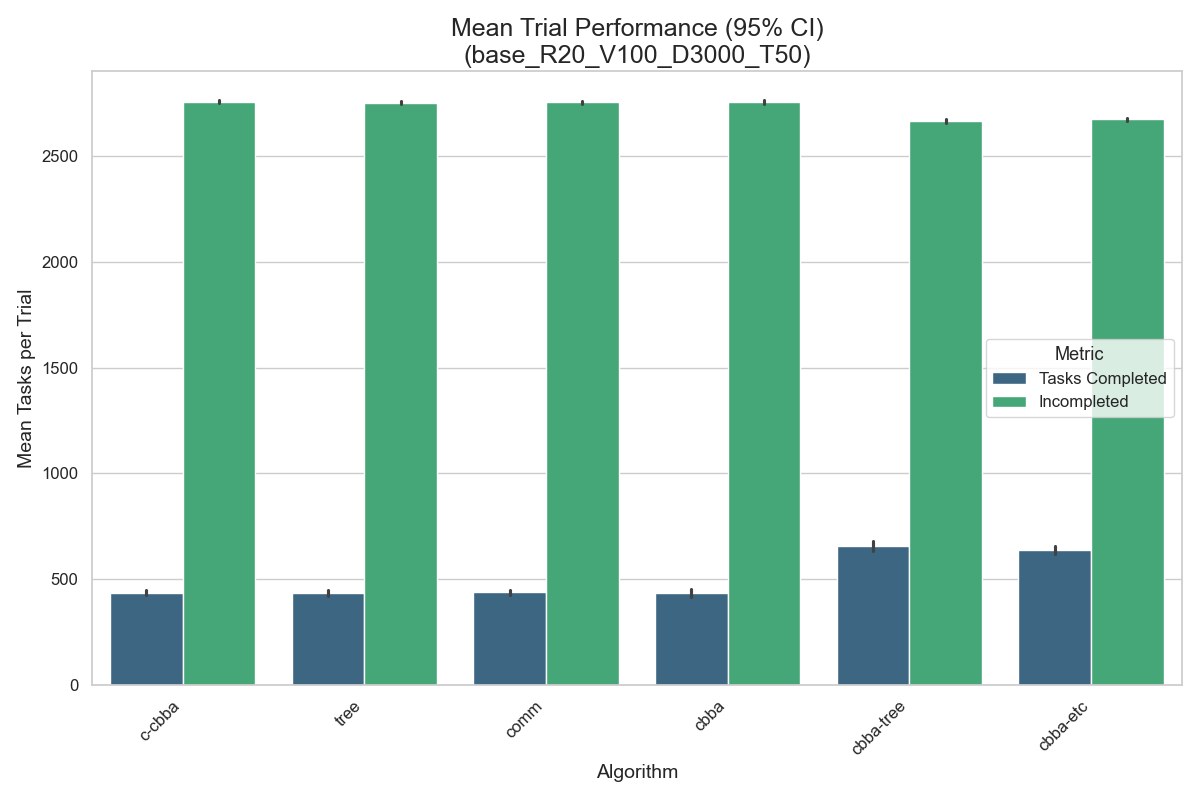}
  \end{center}
\caption{Mean performance of all evaluated algorithms in the baseline scenario, which consisted of 20 robots and 100 active victims over 50 trials. The bars represent the average number of victims rescued versus those lost per trial. The high number of 'Incompleted' tasks reflects victims that expired (100s lifetime) in this challenging steady-state scenario where tasks are continuously replaced. Error bars indicate the 95\% confidence interval. The BT-based consensus algorithms (`cbba-tree`, `cbba-etc`) demonstrate significantly higher effectiveness.
}\label{fig:E01}
\end{figure}

To establish a performance baseline, all algorithms were evaluated under ideal conditions with the previously presented environment called \texttt{base\_R20\_V100}. These conditions were characterized by the absence of action or communication failures, and a steady-state task environment ensured by victim replacement. We used a base environment with 20 robots and 100 victims, where each test was executed for 50 trials. Robot and victim initial positions were randomized. The total aggregated results are presented in Table \ref{tab:key_performance_metrics}, with corresponding per-trial averages and standard deviations displayed in Figure \ref{fig:E01}.

\begin{table}[ht]
\centering
\caption{Summary of key performance metrics for the baseline scenario (\texttt{base\_R20\_V100}). All values represent totals accumulated over 50 trials.}
\label{tab:key_performance_metrics}
\begin{tabular}{lcccc}
\toprule
Algorithm & \begin{tabular}[c]{@{}c@{}}Victims \\ Rescued\end{tabular} & \begin{tabular}[c]{@{}c@{}}Messages \\ Sent\end{tabular} & \begin{tabular}[c]{@{}c@{}}CBBA \\ Negotiations\end{tabular} & \begin{tabular}[c]{@{}c@{}}Failed \\ Rescues\end{tabular} \\
\midrule
\texttt{c-cbba} & 21,856  & 104,535  & 10,687  & 16,547  \\
\texttt{tree} & 21,725  & 0  & N/A  & 43,634  \\
\texttt{comm} & 21,930  & 435,457  & N/A  & 39,193  \\
\texttt{cbba} & 21,756  & 271,415  & 61,685  & 16,160  \\
\texttt{cbba-tree} & \textbf{32,807}  & 81,153  & 28,275  & 23,103  \\
\texttt{cbba-etc} & \textbf{31,800}  & \textbf{27,853}  & \textbf{4,338}  & 23,091  \\
\bottomrule
\end{tabular}
\end{table}

\begin{figure}
  \begin{center}
    \includegraphics[width=0.7\textwidth]{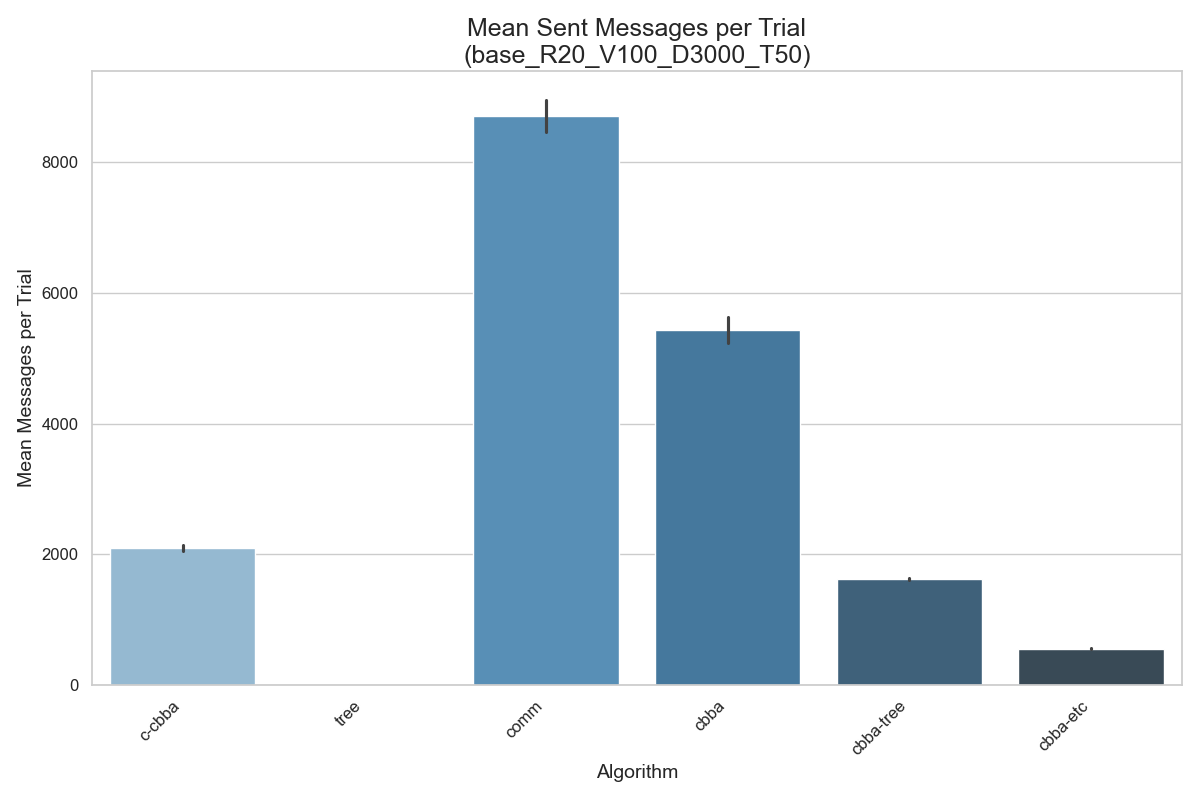}
  \end{center}
  \caption{Comparison of the mean communication overhead for each algorithm in the baseline scenario. The bars show the average number of messages sent per trial. The 'cbba-etc' algorithm demonstrates an important reduction in communication compared to other consensus-based and communicative methods}\label{fig:E02}
\end{figure}

\begin{figure}
  \begin{center}
    \includegraphics[width=0.7\textwidth]{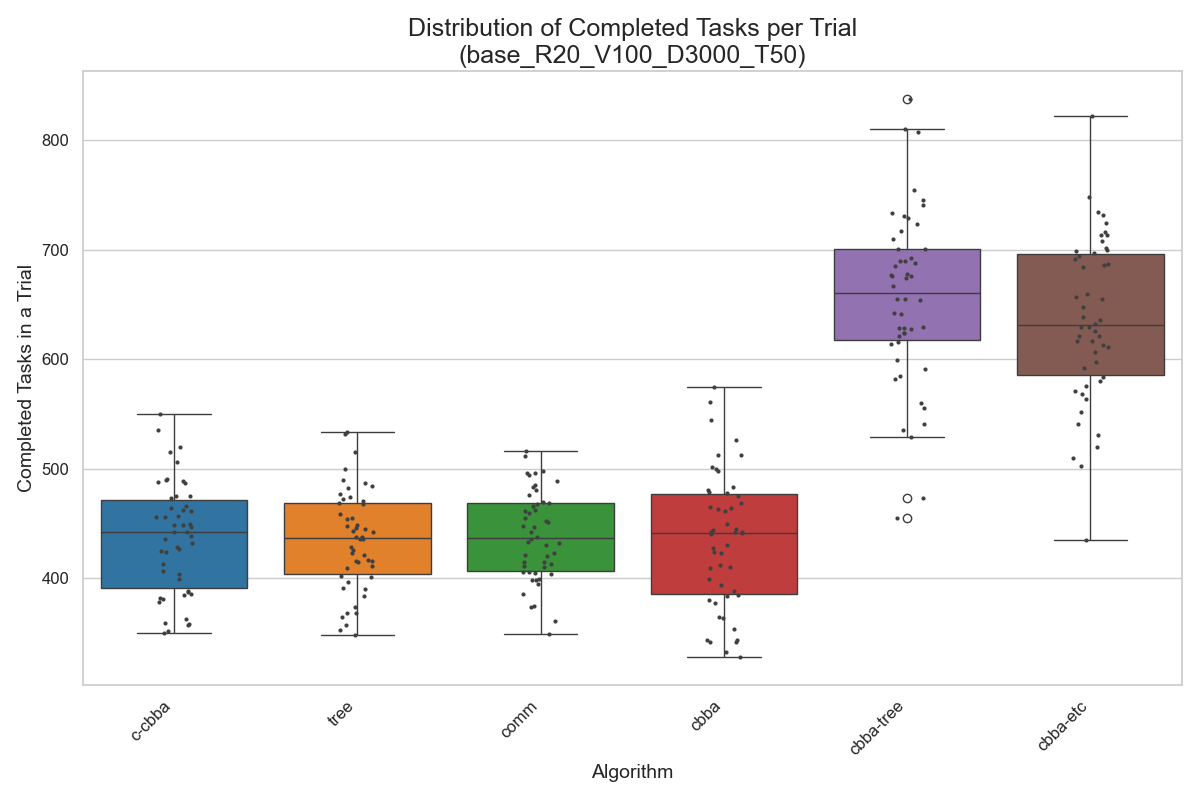}
  \end{center}
  \caption{Distribution of rescued victims per trial for each algorithm in the baseline scenario. The boxplot illustrates the median, quartiles, and range of performance over 50 trials. The results for 'cbba-etc' and 'cbba-tree' show similar effectiveness and are significantly higher than the other algorithms}\label{fig:E03}
\end{figure}

A central challenge in swarm robotics is managing the inherent trade-off between mission effectiveness and the communication overhead required to achieve it. The baseline results, visualized in Figure \ref{fig:E01}, reveal two distinct performance tiers in terms of mission effectiveness. The top tier consists of the BT-based consensus architectures: \texttt{cbba-etc} (31,800 total rescues) and \texttt{cbba-tree} (32,807 total rescues). Statistical analysis (Figure \ref{fig:E03}) confirms their performance is statistically similar ($p \approx 0.324$). The bottom tier comprises all other algorithms: \texttt{c-cbba} (21,856), \texttt{comm} (21,930), \texttt{cbba} (21,756), and the non-communicating \texttt{tree} (21,725). These algorithms perform significantly worse than the top-tier methods ($p < 0.0001$).

However, this effectiveness must be weighed against communication efficiency, shown in Figure \ref{fig:E02}. A significant disparity emerges among the consensus-based methods. \texttt{Cbba-etc} proves to be the most efficient architecture, requiring only \textbf{27,853 messages} to coordinate its actions. This is substantially lower than the periodic \texttt{cbba-tree} (81,153 messages) and the \texttt{c-cbba} baseline (104,535 messages). The reactive \texttt{cbba} algorithm was the least efficient, generating 271,415 messages.

Figure \ref{fig:E05} provides a clear cost-benefit analysis. These results shows that CBBA-ETC provides a good solution to this classic performance dilemma: it achieves the maximum effectiveness (top-tier) with the minimum communication cost (highest efficiency). Notably, the state-of-the-art \texttt{c-cbba} algorithm is not only 3.7 times less efficient (104,535 vs 27,853 messages) but also achieves 31\% fewer rescues in this scenario (21,856 vs 31,800).

\subsection{Scalability Analysis}

To evaluate how the coordination architectures respond to increasing environmental complexity and dynamism, we analyzed their scalability with respect to task density. This involved varying the number of simultaneously available victims (25, 50, 100, 200, and 500) while keeping the number of robots fixed at 20 (\texttt{base\_R20\_V\{25, 50, 100, 200, 500\}}). This analysis is crucial for understanding performance in scenarios ranging from sparse task distributions to highly saturated environments, testing the adaptive capabilities of algorithms like CBBA-ETC.
  
\subsubsection{Effectiveness under Varying Task Loads}

\begin{figure}
  \begin{center}
    \includegraphics[width=0.95\textwidth]{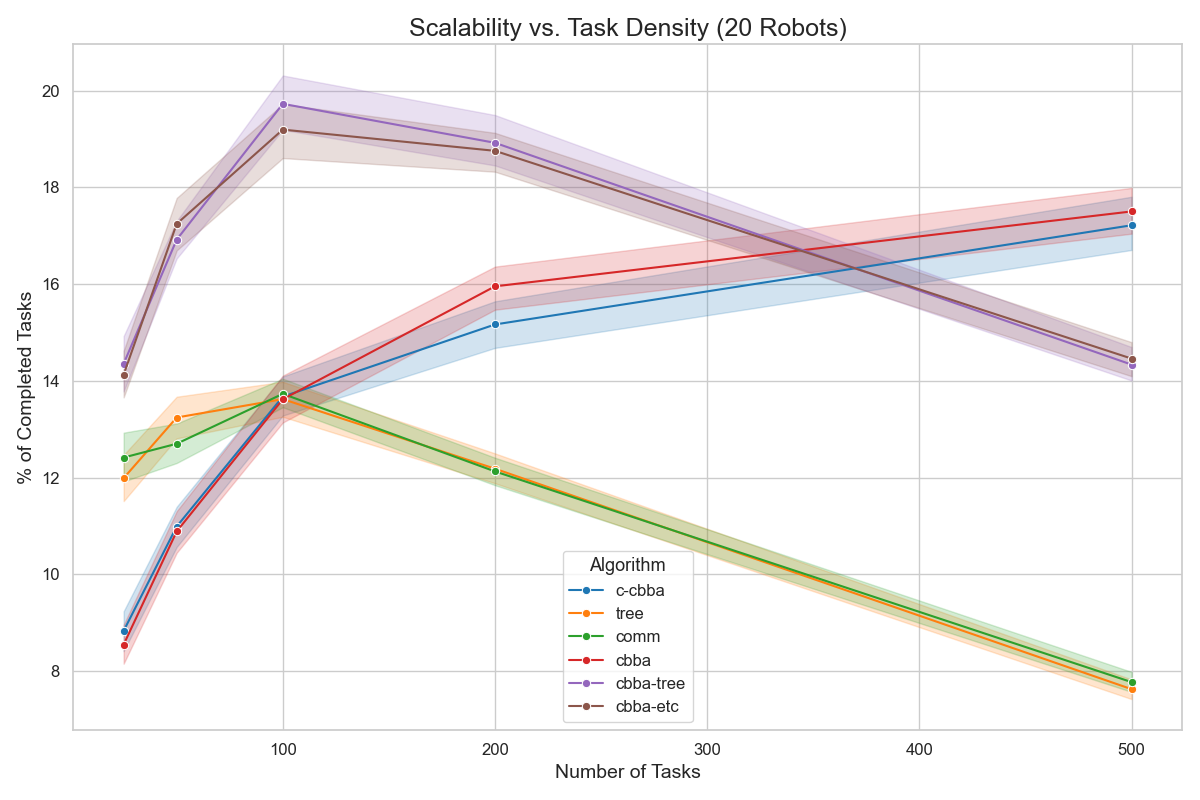}
  \end{center}
  \caption{Scalability of the algorithms with respect to task density, showing the percentage of rescued victims as the number of initial victims increases from 25 to 500, with a fixed team of 20 robots. The consensus-based algorithms that use Behavior Trees ('cbba-etc' and 'cbba-tree') maintain the highest performance up to moderate saturation (V200). At high saturation (V500), 'cbba' and 'c-cbba' achieve slightly higher absolute rescue counts, but Figure \ref{fig:E062} reveals this comes at a significant communication cost.}\label{fig:E06}
\end{figure}

Figure \ref{fig:E06} illustrates the mission effectiveness (percentage of rescued victims) as task density increases. At lower densities (V25, V50, V100), the BT-based consensus algorithms, \texttt{cbba-etc} and \texttt{cbba-tree}, consistently demonstrate superior performance, forming the top tier. For instance, at V100, they achieve rescue rates around 19-20\%, significantly higher than the other algorithms which cluster around 13-14\%.

However, a shift occurs in the high-saturation scenario (V500). Here, \texttt{cbba} (142,898 rescues)  and \texttt{c-cbba} (140,368 rescues)  achieve the highest absolute number of rescues, marginally surpassing \texttt{cbba-etc} (115,887 rescues)  and \texttt{cbba-tree} (114,751 rescues). This suggests that under extreme task saturation, the very high communication frequency of \texttt{cbba} and the structured approach of \texttt{c-cbba} allow them to take better advantage of the abundance of tasks, achieving a slightly higher total volume of completed work.

\subsubsection{Efficiency Across Task Densities}

\begin{figure}
  \begin{center}
    \includegraphics[width=0.95\textwidth]{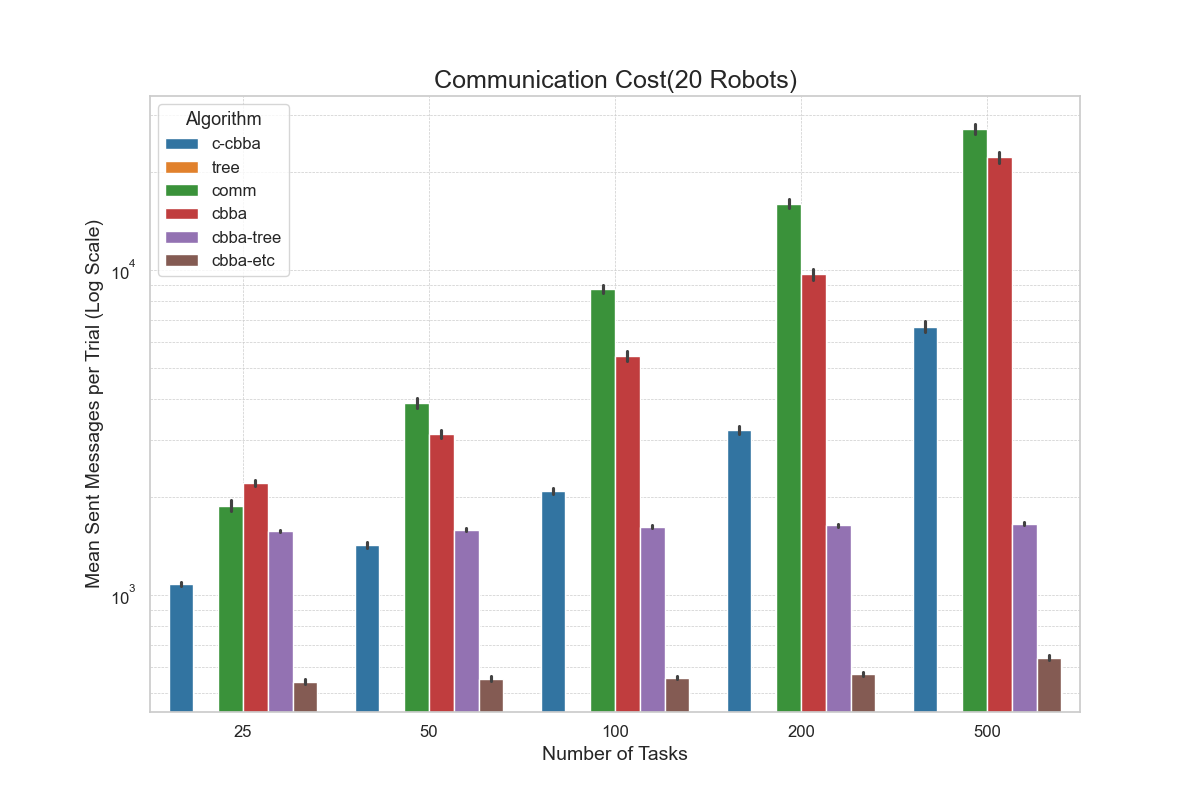}
  \end{center}
  \caption{Communication cost as victim density increases (logarithmic scale) for a fixed team of 20 robots. While 'cbba' and 'comm' exhibit exponential increases in messages, and 'c-cbba' shows significant scaling, 'cbba-etc' maintains high performance with only a marginal increase in communication, demonstrating superior efficiency at scale.}\label{fig:E062}
\end{figure}

While effectiveness under saturation is significant, the communication cost reveals the practical limitations. Figure \ref{fig:E062} plots the mean number of messages sent per trial (logarithmic scale) against task density. The key finding is the prohibitive cost associated with the top performers in the V500 scenario. \texttt{cbba} achieves its high rescue count at the cost of network explosion, generating approximately 1.11 million messages. Similarly, \texttt{c-cbba}'s performance requires 334,648 messages. In contrast, \texttt{cbba-etc} maintains remarkable efficiency even under extreme load, requiring only 32,023 messages, scaling marginally from its baseline communication level. The periodic \texttt{cbba-tree} also remains relatively efficient, using 82,862 messages.

\subsubsection{Conclusion on Task Density Scalability}

The analysis demonstrates CBBA-ETC's superior efficiency in scaling with task density. While \texttt{cbba} and \texttt{c-cbba} achieve approximately 21-23\% more rescues under extreme saturation (V500), they do so at communication costs that are roughly 34 times and 10 times higher, respectively, compared to CBBA-ETC. Such high network traffic is often unsustainable in resource-constrained environments. CBBA-ETC offers a compelling balance, delivering performance close to the maximum observed effectiveness but with a significantly lower and more sustainable communication overhead. This validates the effectiveness of its event-triggered and adaptive consensus mechanisms, enabling self-regulation and maintaining efficiency even when the environment becomes highly dynamic and task-saturated.

\subsection{Scalability Analysis (Robot Density)}

To evaluate the scalability of the coordination algorithms with respect to the size of the swarm, we conducted experiments varying the number of robots (5, 10, 20, and 40) while keeping the number of tasks constant at 100 (\texttt{base\_R\{5, 10, 20, 40\}\_V100}). This analysis assesses how the coordination mechanisms handle increased potential for interaction and conflict as more agents operate in the same environment.

\subsubsection{Effectiveness Across Swarm Sizes}

\begin{figure}
  \begin{center}
    \includegraphics[width=0.95\textwidth]{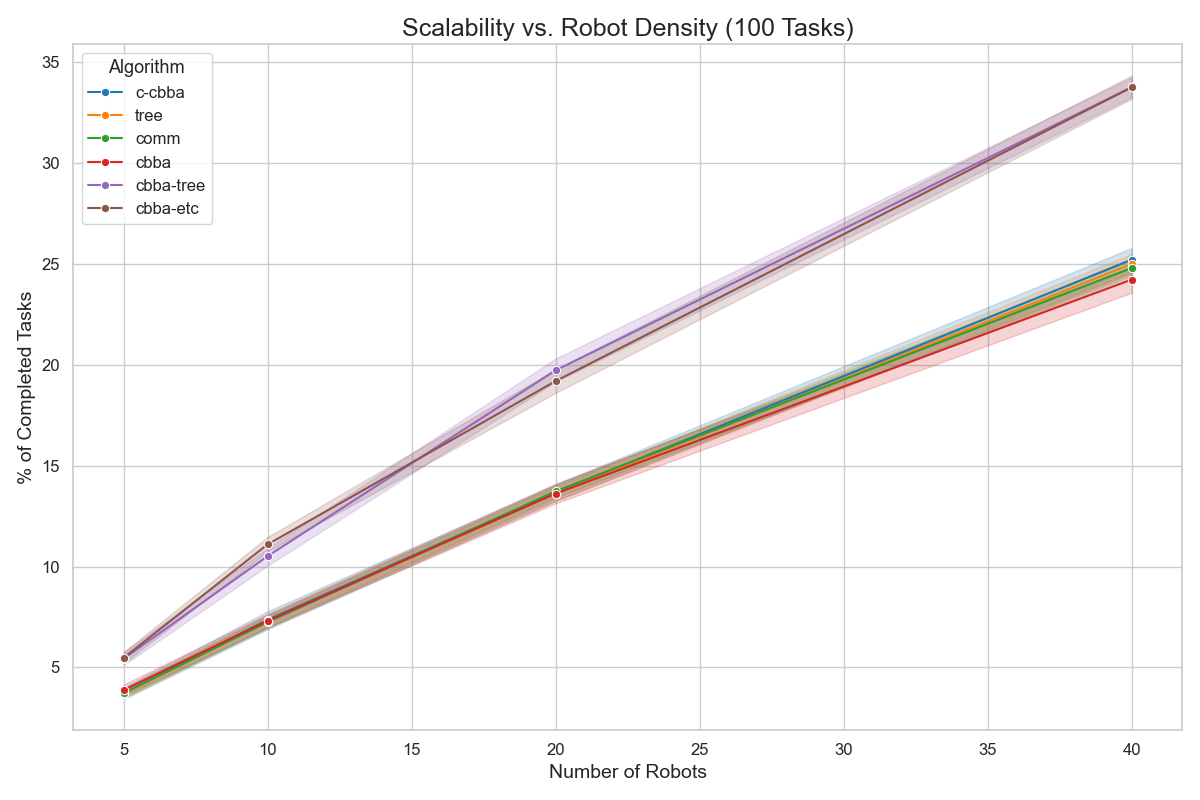}
  \end{center}
  \caption{Scalability of the algorithms with respect to robot density, showing the percentage of rescued victims as the number of robots increases from 5 to 40, with a fixed set of 100 victims. The consensus-based algorithms that use Behavior Trees ('cbba-etc' and 'cbba-tree') maintain the highest performance across all swarm sizes, while 'c-cbba' consistently performs in the lower tier.}\label{fig:E063}
\end{figure}

Figure \ref{fig:E063} plots the mission effectiveness (percentage of rescued victims) as the number of robots increases. The results clearly show that the two distinct performance tiers observed in the baseline scenario remain consistent across all swarm sizes. The BT-based consensus architectures, \texttt{cbba-etc} and \texttt{cbba-tree}, consistently occupy the top tier, achieving significantly higher rescue rates compared to the other algorithms. For example, with 40 robots, \texttt{cbba-etc} and \texttt{cbba-tree} rescue approximately 33.8\% of victims each, maintaining their lead.

The bottom tier consistently includes \texttt{c-cbba}, \texttt{cbba}, \texttt{comm}, and \texttt{tree}. Even with 40 robots, their performance clusters around 24-25\%, significantly below the top-tier algorithms. This demonstrates that the architectural advantages of integrating CBBA with Behavior Trees provide a scalable benefit in mission effectiveness as the swarm size increases. The state-of-the-art \texttt{c-cbba} remains in the lower performance tier across all tested robot densities.

\subsubsection{Communication Efficiency with Increasing Robots}

While effectiveness scales positively for the top-tier algorithms, maintaining communication efficiency is crucial as more agents join the network. Our analysis confirms that \texttt{cbba-etc} remains the most communication-efficient consensus algorithm across all swarm sizes tested. The advantage becomes particularly pronounced in larger swarms. With 40 robots, \texttt{cbba-etc} required only 115,855 messages  to achieve its top-tier performance. In contrast:

\begin{itemize}

  \item c-cbba needed 226,441 messages  (approximately 2 times more than \texttt{cbba-etc}) while achieving significantly lower effectiveness.
  \item The periodic \texttt{cbba-tree} used 329,545 messages  (nearly 3 times more).
  \item The reactive \texttt{cbba} generated over 1.05 million messages  (approximately 9 times more), highlighting the inefficiency of frequent, untargeted communication in larger groups.

\end{itemize}

\subsubsection{Conclusion on Robot Density Scalability}

The results demonstrate that the benefits of the CBBA-ETC architecture scale effectively with the number of robots. It consistently maintains its position in the top tier of mission effectiveness alongside \texttt{cbba-tree}, significantly outperforming \texttt{c-cbba}, \texttt{cbba}, and simpler strategies. Crucially, it achieves this high effectiveness while remaining the most communication-efficient consensus algorithm, with its advantage in message reduction becoming even more significant as the swarm size increases. This validates CBBA-ETC as a scalable solution for coordinating larger multi-robot teams in dynamic environments.

\subsection{Robustness Analysis}

A critical requirement for multi-robot systems, particularly those operating in unpredictable  environments, is resilience to various forms of operational failure. Given the resource-constrained nature of real-world deployments and the inherent unreliability of wireless channels, maintaining effective coordination under communication degradation is paramount. This section rigorously evaluates the robustness of the compared algorithms against communication degradation (packet loss and bandwidth limitations), physical action failures, and permanent agent loss, highlighting how the synergistic integration of Event-Triggered Control (ETC) and Behavior Trees (BTs) in the proposed CBBA-ETC architecture provides a critical, communication-aware contribution to system resilience.

\subsubsection{Robustness Against Action Execution Failures}

To assess resilience to physical uncertainty, we introduced stochastic failures for the robots' core actions (move, inspect, rescue). We compared the baseline performance (0\% failure) against scenarios with 25\% and 50\% failure probability for each action.

\begin{figure}[h!]
  \begin{center}
    \includegraphics[width=0.95\textwidth]{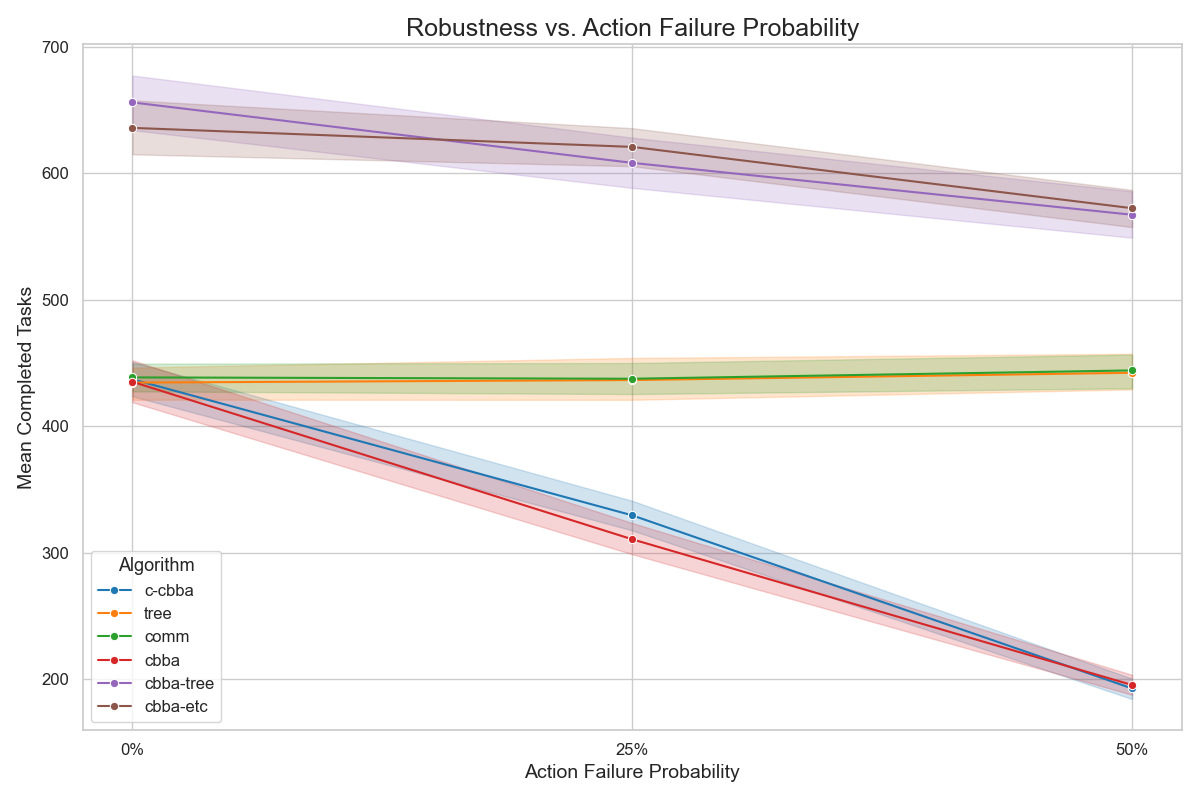}
  \end{center}
  \caption{Performance degradation of each algorithm as the probability of action failure increases from 0\% to 50\%. The lines plot the mean number of rescued victims, with shaded areas representing the 95\% confidence interval. Critically, the non-BT architectures ('c-cbba', 'cbba') collapse at high failure rates, while all BT-based architectures demonstrate significant robustness.}\label{fig:E04}
\end{figure}

Figure \ref{fig:E04} illustrates the performance degradation as action failure probability increases. A critical finding emerges at the 50\% failure rate: the performance of algorithms \textit{not} employing a Behavior Tree for execution control collapses. Specifically, \texttt{cbba} manages only 9,763 rescues , and \texttt{c-cbba} achieves just 9,627 rescues, representing a drastic drop from their baseline performance.

In contrast, all architectures based on Behavior Trees (BTs) demonstrate remarkable robustness. Even under a 50\% action failure rate, \texttt{cbba-etc} (28,623 rescues), \texttt{cbba-tree} (28,364 rescues), \texttt{comm} (22,207 rescues), and \texttt{tree} (22,113 rescues) maintain a significantly higher level of effectiveness.

This result strongly suggests that the observed robustness against physical failures primarily arises from the Behavior Tree execution architecture, rather than the consensus algorithm itself. The BT provides inherent mechanisms for handling local contingencies (e.g., retrying failed actions, executing fallback behaviors) without requiring immediate strategic re-coordination . This validates the third contribution claimed in this paper regarding enhanced resilience through modular execution. Within the robust BT-based group, \texttt{cbba-etc} and \texttt{cbba-tree} remain the optimal architectures, successfully combining the resilience of the BT framework with the coordination effectiveness of the CBBA consensus protocol.

\subsubsection{Robustness Against Communication Degradation and Agent Loss}

\begin{figure}[h!]
  \begin{center}
    \includegraphics[width=0.95\textwidth]{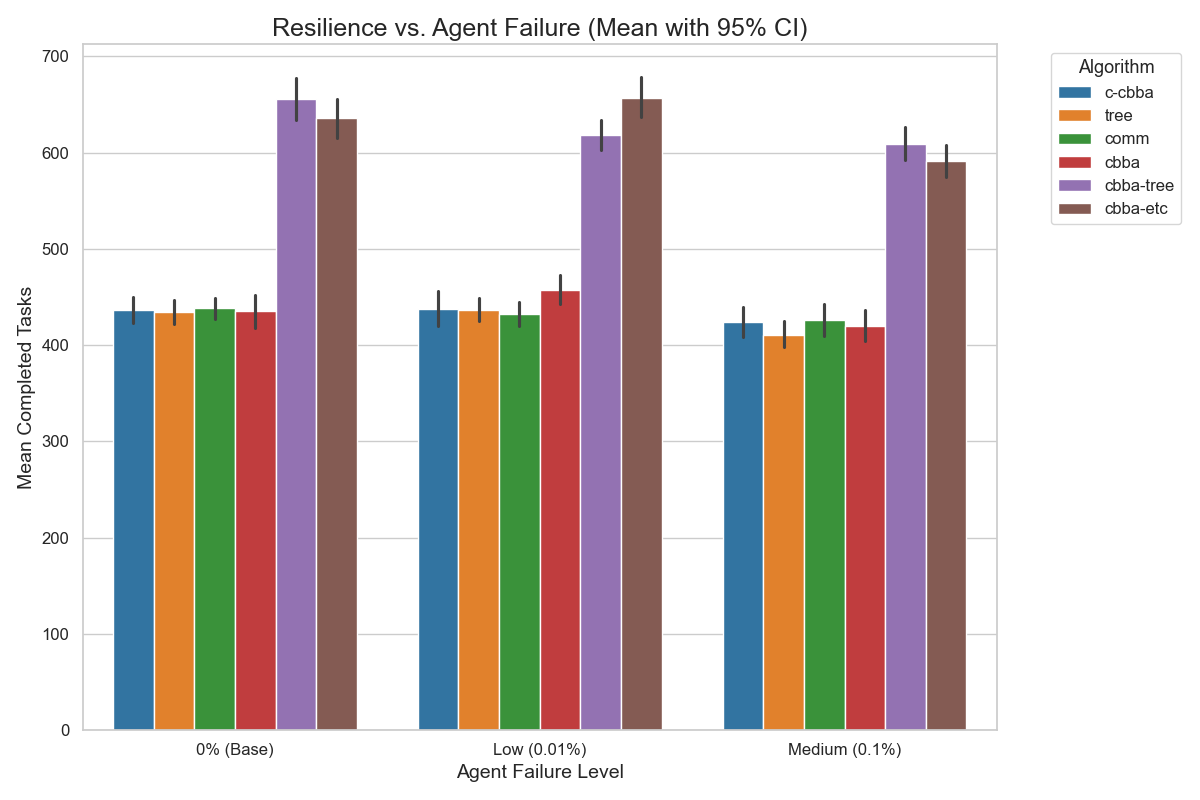}
  \end{center}
  \caption{System resilience to permanent agent failure, showing the mean number of rescued victims as the probability of agent loss per step increases. The performance of all decentralized algorithms degrades gracefully. 'cbba-etc' and 'cbba-tree' maintain their high-performance advantage even as the number of active robots in the swarm decreases over time.}\label{fig:E08}
\end{figure}

\begin{figure}[h!]
  \begin{center}
    \includegraphics[width=0.95\textwidth]{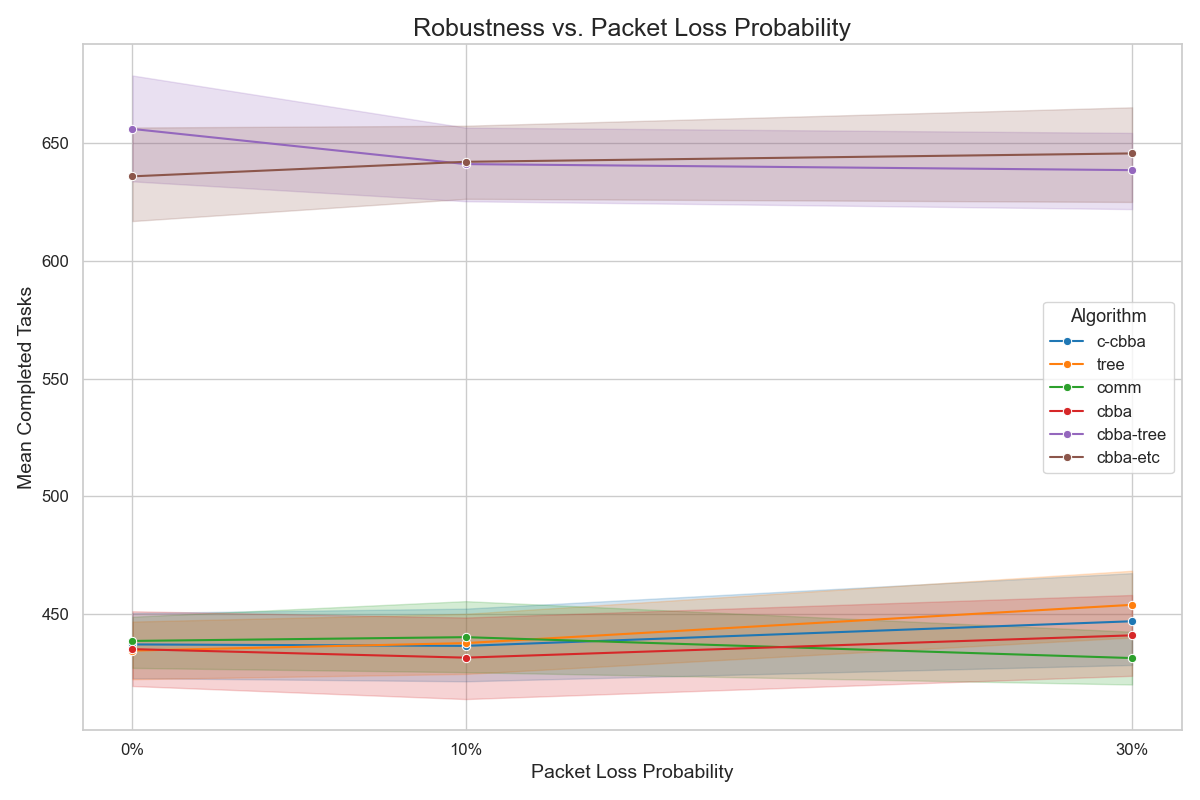}
  \end{center}
  \caption{Algorithm robustness to communication degradation, plotting the mean number of rescued victims against increasing packet loss probability (with bandwidth limits). The BT-based consensus architectures ('cbba-etc' and 'cbba-tree') consistently outperform other methods across all tested levels of communication failure, demonstrating significant resilience.}\label{fig:E09}
\end{figure}

We further evaluated resilience against network imperfections and agent losses using scenarios with communication packet loss (10\% and 30\%, with bandwidth limits) and permanent agent failure (0.01\% and 0.1\% probability per step after 500s). Figures \ref{fig:E09} and \ref{fig:E08} illustrate the performance under these conditions. The key observation is that the two performance tiers identified in the baseline analysis are consistently maintained across all these failure scenarios.

\begin{itemize}
  \item Communication Packet Loss (Figure \ref{fig:E09}): Even with 30\% packet loss, \texttt{cbba-etc} (32,287 rescues ) and \texttt{cbba-tree} (31,932 rescues ) remain the top performers. \texttt{c-cbba} (22,348 rescues ) and the other algorithms continue to operate at a significantly lower effectiveness level.
  \item Permanent Agent Failure (Figure \ref{fig:E08}): Similarly, with a medium agent failure rate (0.1\%), \texttt{cbba-etc} (29,556 rescues ) and \texttt{cbba-tree} (30,476 rescues ) maintain their superior performance compared to \texttt{c-cbba} (21,219 rescues ) and the remaining algorithms. The decentralized nature allows graceful degradation as the number of active agents decreases.
\end{itemize}

\subsubsection{Conclusion on Robustness}

The CBBA-ETC architecture demonstrates significant robustness against various operational failures. Its resilience to physical action failures is primarily attributed to the integrated Behavior Tree framework, validating its importance for reliable execution in uncertain environments. Furthermore, CBBA-ETC maintains its top-tier effectiveness relative to baselines even under conditions of considerable communication packet loss and progressive agent attrition. This comprehensive resilience profile underscores its suitability for deployment in challenging real-world scenarios like Search and Rescue.

\subsection{Analysis of Heterogeneity Management Efficiency (Finite Tasks)}

\begin{figure}[h!]
  \begin{center}
    \includegraphics[width=0.95\textwidth]{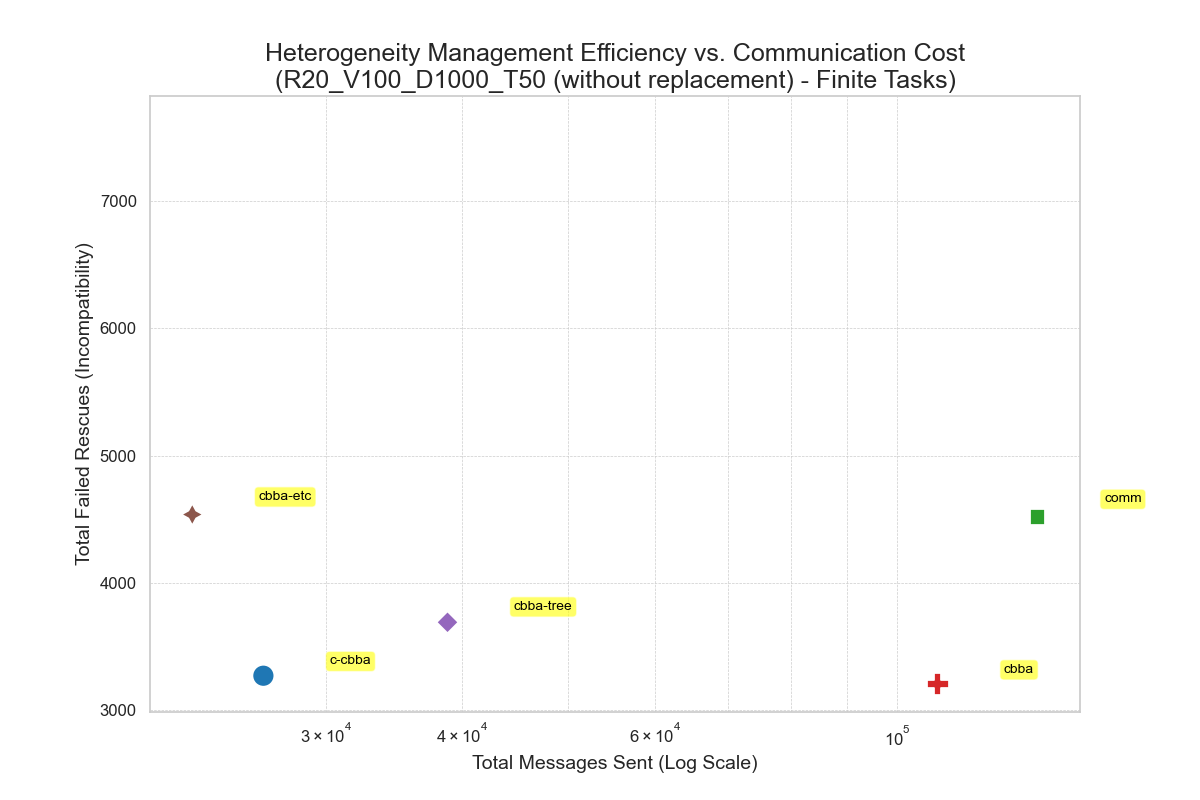}
  \end{center}
  \caption{Analysis of heterogeneity management efficiency versus communication cost in the finite-task scenario. The plot shows Total Failed Rescues (due to capability mismatch) against Total Messages Sent (logarithmic scale). The ideal performance region is the bottom-left corner (minimal failed rescues and minimal communication). \texttt{Cbba-etc} and \texttt{c-cbba} demonstrate the best overall balance, achieving low communication cost while effectively managing heterogeneity, positioned closer to the ideal region compared to the communication-heavy \texttt{cbba} and \texttt{comm} algorithms. For visualization purposes on the logarithmic message axis, the non-communicating tree algorithm is not shown. It performed worst in managing heterogeneity, accumulating 7,607 failed rescues because it cannot share or receive information about task requirements.}\label{fig:E20}
\end{figure}

To further analyze performance in a scenario pertinent to SAR operations where a specific set of tasks must be completed, we evaluated the algorithms in a finite-task environment. In this setup, 100 initial victims were present, and they were not replaced upon rescue, shifting the focus from steady-state throughput to the efficiency of completing a fixed workload, particularly in managing team heterogeneity. The simulation duration was set to 1000s. Results are presented in figure \ref{fig:E20}.

\subsubsection{Effectiveness in Task Completion}

In this finite-task context, the total number of rescued victims is less indicative of performance differences, as most algorithms eventually manage to rescue nearly all available victims given sufficient time. The primary challenge shifts from maximizing throughput to minimizing wasted effort, especially concerning the heterogeneity constraint (color matching).

\subsubsection{Efficiency in Managing Heterogeneity (Failed Rescues)}

The metric 'Failed Rescues' becomes crucial here, quantifying the number of times robots approached and inspected victims only to find their capabilities (color) did not match the task requirement. This represents wasted time and energy.

\begin{itemize}
\item  The non-communicating \texttt{tree} algorithm performed the worst, accumulating 7,607 failed rescues due to its inability to share or receive information about task requirements or assignments.
  \item  Algorithms with high communication frequency, \texttt{cbba} (3,206 fails) and \texttt{c-cbba} (3,271 fails), were the most effective at minimizing these failures. Their frequent message exchanges likely allow information about discovered victim colors and assignments to propagate more rapidly through the swarm, preventing incompatible robots from pursuing those tasks.
  \item  The BT-based consensus algorithms demonstrated a strong balance. \texttt{cbba-tree} recorded 3,691 failed rescues, while \texttt{cbba-etc} recorded 4,538. Both significantly outperform the \texttt{tree} baseline, indicating effective management of heterogeneity, albeit with slightly more failed attempts than the communication-heavy methods. The simple \texttt{comm} algorithm also performed reasonably well with 4,520 fails.
  \end{itemize}

\subsubsection{Mission Cost (Communication Overhead)}

Analyzing the communication cost is essential for evaluating overall efficiency in this scenario:

\begin{itemize}
\item The algorithms that minimized failed rescues did so at a high communication cost: \texttt{cbba} sent 108,920 messages, and \texttt{comm} sent 134,496 messages (the highest).
\item In contrast, \texttt{cbba-etc} and \texttt{c-cbba} proved to be by far the most efficient. \texttt{cbba-etc} required only 22,636 messages, the lowest among communicating algorithms, while \texttt{c-cbba} used 26,296 messages. \texttt{cbba-tree} had a moderate cost of 38,764 messages.
\end{itemize}

\subsubsection{Conclusion on Heterogeneity Efficiency}

In the finite-task scenario, where completing a set workload efficiently is paramount, \texttt{cbba-etc} and \texttt{c-cbba} offer the best overall balance between effective heterogeneity management and resource conservation. While \texttt{c-cbba} (along with the inefficient \texttt{cbba}) demonstrates a marginal advantage in minimizing failed rescue attempts, \texttt{cbba-etc} achieves excellent heterogeneity management performance while maintaining its position as the most communication-efficient architecture. This highlights its ability to make effective trade-offs, ensuring robust coordination and good resource utilization even when communication is selectively triggered.

\subsection{Performance Synthesis of the CBBA-ETC Framework}\label{adaptability-robustness-and-efficiency-of-the-cbba-etc-expert-system}
 
The presented experimental results collectively demonstrate that the CBBA-ETC framework provides a highly effective and resource-efficient solution to the multi-robot task allocation problem. This architecture excels in adaptability, robustness, and, most critically, communication efficiency. When compared against the selected baselines—including purely reactive (\texttt{tree}), simple communication (\texttt{comm}), reactive CBBA (\texttt{cbba}), periodic CBBA integrated with BTs (\texttt{cbba-tree}), and the state-of-the-art Clustering-CBBA (\texttt{c-cbba}), the strengths of the CBBA-ETC design become evident across a wide range of operational challenges.
  
Consistently across baseline conditions and varying robot densities (5 to 40 robots), CBBA-ETC achieves top-tier mission effectiveness, measured by the number of rescued victims. Its performance in this regard is statistically comparable to the periodic \texttt{cbba-tree}, and significantly superior to \texttt{c-cbba}, \texttt{cbba}, \texttt{comm}, and \texttt{tree}. As illustrated in Figure \ref{fig:E063}, these two distinct performance tiers persist regardless of swarm size. For instance, with 40 robots, \texttt{cbba-etc} and \texttt{cbba-tree} rescue approximately 33.8\% of victims each, while the remaining algorithms cluster around 24-25\%.

\begin{figure}
  \begin{center}
    \includegraphics[width=0.95\textwidth]{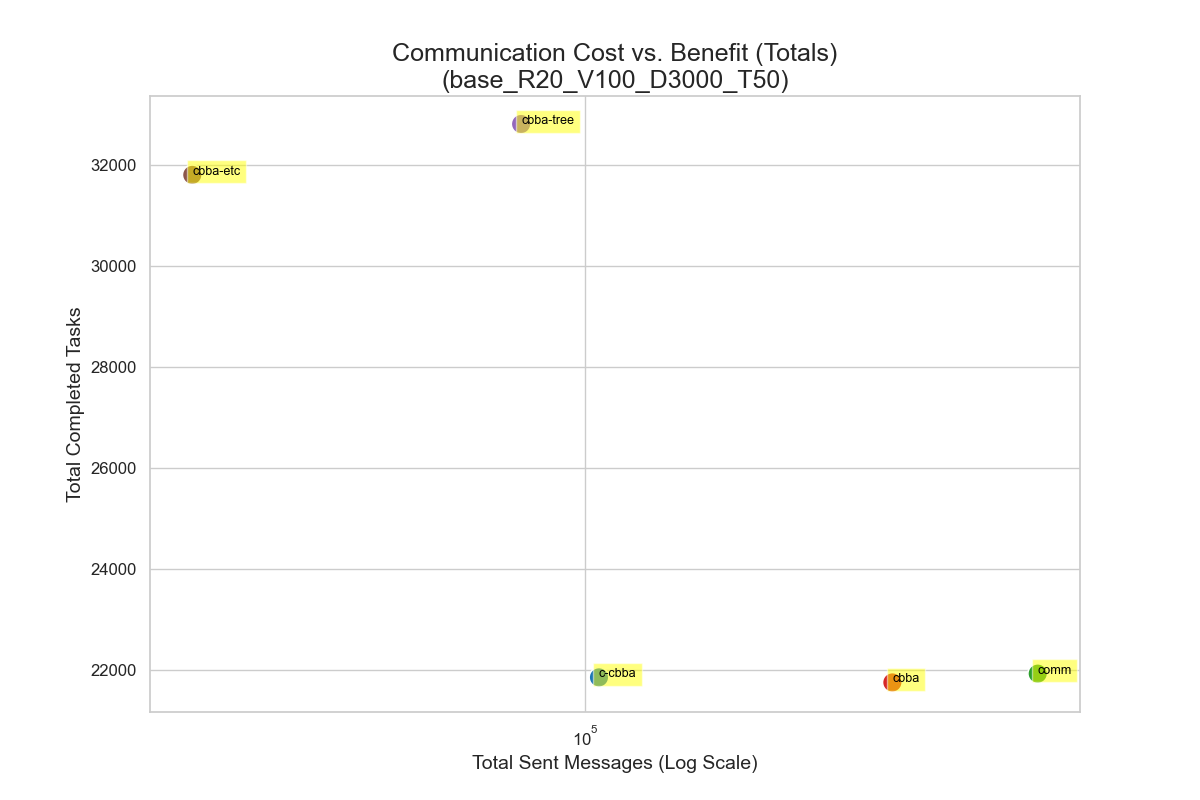}
  \end{center}
  \caption{A cost-benefit analysis of the algorithms under baseline conditions, plotting total rescued victims against the total number of messages sent (on a logarithmic scale). This visualization highlights the trade-off between mission effectiveness and communication efficiency. 'Cbba-etc' is positioned in the upper-left, demonstrating its performance with minimal communication cost.}\label{fig:E05}
\end{figure}

However, CBBA-ETC distinguishes itself significantly through its communication efficiency. Across all scenarios, it consistently utilizes a fraction of the network resources required by other consensus-based methods. This advantage becomes more pronounced under demanding conditions. In the high task-density scenario (V500, Figure \ref{fig:E062}), while \texttt{cbba} and \texttt{c-cbba} achieved slightly higher absolute rescue counts (approx. 21-23\% more than CBBA-ETC), they did so at a prohibitive communication cost, over 1.11 million messages for \texttt{cbba} (34x more than CBBA-ETC) and 334,648 for \texttt{c-cbba} (10x more). CBBA-ETC maintained high effectiveness with only 32,023 messages. Similarly, with 40 robots, CBBA-ETC used only 115,855 messages, roughly half that of \texttt{c-cbba} (226,441) and nine times less than \texttt{cbba} (1.05M). This highlights the superior scalability and sustainability of the event-triggered approach (Figure \ref{fig:E05}).

Furthermore, the framework demonstrates significant robustness. As shown in Figure \ref{fig:E04}, the integration of Behavior Trees provides substantial resilience against physical action failures. While the performance of non-BT architectures like \texttt{cbba} and \texttt{c-cbba} collapsed under a 50\% failure rate (rescuing fewer than 10,000 victims), all BT-based systems, including CBBA-ETC (28,623 rescues), maintained much higher effectiveness. CBBA-ETC also maintains its relative performance advantage under communication degradation (Figure \ref{fig:E09}) and permanent agent loss (Figure \ref{fig:E08}), where the two performance tiers persist even with 30\% packet loss or a 0.1\% agent failure probability per step.

Ultimately, CBBA-ETC consistently delivers top-tier mission effectiveness comparable to the best-performing periodic strategy (\texttt{cbba-tree}), but achieves this with substantially lower communication overhead across varying scales and under diverse failure conditions. This efficiency is not merely a secondary benefit; it is crucial for practical deployment, implying lower energy consumption, reduced network congestion, and potentially faster decision cycles. In synthesizing these results, the CBBA-ETC architecture stands out by effectively balancing high performance, remarkable efficiency, and robust operation, positioning it as a highly suitable model for coordinating multi-robot teams in demanding, dynamic, and resource-constrained environments such as SAR operations.

\subsection{\texorpdfstring{\textbf{Limitations and Trade-offs}}{Limitations and Trade-offs}}
\label{limitations-and-trade-offs}

While the \texttt{cbba-etc} framework demonstrates a superior combination of performance and efficiency across a majority of the tested scenarios, the comprehensive analysis also reveals important limitations and trade-offs. We acknowledge that our study is based on several simplifying assumptions designed to isolate core coordination challenges. Understanding these limitations and the specific contexts where \texttt{cbba-etc}'s advantages diminish is crucial for its practical application.

\subsubsection{Balancing Reactivity and Stability}
The primary trade-off identified is a classic engineering challenge: balancing reactive efficiency against periodic stability. This is most evident in environments with extremely high physical uncertainty. In the scenario analyzed in Figure \ref{fig:E04} with a 50\% action failure rate, the time-based \texttt{cbba-tree} outperformed \texttt{cbba-etc} in the total number of rescues. This suggests that when the environment becomes excessively "noisy" due to unreliable low-level actions, the stable, metronomic cadence of \texttt{cbba-tree}'s periodic consensus is more robust. The highly reactive nature of \texttt{cbba-etc}, while beneficial in most cases, may be prompted into performing less productive re-negotiations in response to transient execution failures. This presents a clear choice for a mission planner: one must weigh the superior stability of periodic consensus against the conserved resources offered by an event-triggered model in highly unpredictable environments.

\subsubsection{Simplifications of the Physical Model and Validation}
Our simulation intentionally abstracts certain physical complexities to focus on the algorithmic challenges of coordination.
\begin{itemize}
    \item \textbf{Obstacles and Path Planning:} Our model intentionally abstracts low-level navigation by not including physical obstacles. This is a deliberate design choice to isolate and rigorously evaluate the core challenge of decentralized task allocation, which is particularly relevant for our primary scenario of UAVs operating in open airspace where pathfinding is often trivial. However, the framework is designed for extensibility. To apply it to more complex environments, such as ground robots in cluttered areas, its modular architecture allows for the straightforward integration of a path planner (e.g., A*). The utility function, $U_i(j)$, would simply be updated to use the actual path distance from the planner instead of the Euclidean distance, leaving the core consensus logic unaltered. This demonstrates the model's flexibility to be adapted for more complex navigational challenges.
    \item \textbf{Constant Velocity:} Our model assumes a constant maximum speed, which is a reasonable simplification for the primary scenario of UAVs (drones) navigating via GPS in open airspace, where cruising speed is relatively stable. However, to extend the framework to more complex situations—such as for Unmanned Ground Vehicles (UGVs) traversing varied terrain or for UAVs operating in strong winds—this assumption would need to be adapted. The architecture can readily accommodate this complexity by shifting the utility metric from pure distance to Estimated Time of Arrival (ETA). This would allow the system to naturally account for speed variations due to terrain, battery levels, or other dynamic factors, enhancing its applicability to heterogeneous teams and more challenging environments.
    \item \textbf{Energy Consumption Model:} Furthermore, we acknowledge that our simulation model abstracts away certain physical indicators, most notably a detailed agent energy consumption model. While this is a deliberate simplification to isolate the coordination dynamics, the communication overhead serves as a strong proxy for the energy expenditure related to inter-agent coordination. In real-world robotic platforms, wireless data transmission is a significant source of power drain. Therefore, the order-of-magnitude reduction in network traffic demonstrated by CBBA-ETC not only alleviates network congestion but also strongly implies a corresponding improvement in energy efficiency and operational endurance for the swarm. Quantifying this energy saving on physical hardware is a critical objective for our future work.
    \item \textbf{Simulation-Only Validation:} The evaluation was conducted exclusively in simulation. Real-world environments present additional complexities that have not been modeled. The validation on physical robot platforms is the crucial next step of our research, as detailed in the Future Work section.
\end{itemize}

\subsubsection{Communication Robustness}
A further limitation is \texttt{cbba-etc}'s relative sensitivity to high rates of communication failure, as shown in Figure \ref{fig:E09}. Its efficiency is derived from transmitting fewer, more critical messages. The drawback is that the loss of these messages has a more significant impact than in a system with redundant, periodic communication. However, the fact that it maintains top-tier performance even while losing nearly a third of its messages can also be interpreted as a remarkable demonstration of the system's resilience.

\subsubsection{Parameter Tuning}
Finally, a notable limitation is the absence of a formal parametric study to analyze the system's sensitivity to its core parameters, such as the event-triggering thresholds ($\theta$) and the adaptive interval factors ($\kappa, \lambda$). While the parameters used in this study were the result of a rigorous empirical tuning process aimed at achieving robust performance across our test scenarios, we acknowledge that their optimal values may be environment-dependent. A comprehensive sensitivity analysis was beyond the scope of this work, but we identify the development of mechanisms for the system to autonomously learn or adapt these parameters online as a critical direction for future research, which would further enhance its adaptability and reduce the need for manual calibration.

\section{Conclusions and Future Work\label{section:Conclusions}}

Coordinating multi-robot systems (MRS) effectively under the communication constraints inherent in dynamic environments like Search and Rescue (SAR) remains a significant challenge. This paper introduced and validated a novel framework for event-triggered organization, CBBA-ETC, designed to enable highly efficient and adaptive task allocation within heterogeneous robotic swarms operating over resource-limited networks. Our approach leverages an adaptive consensus mechanism where network communication for task negotiation is strategically initiated only in response to significant events, coupled with swarm-level self-regulation of coordination pace and robust individual agent execution.

The core contribution, demonstrated through extensive comparative simulations, is the framework's  network resource efficiency. CBBA-ETC drastically reduces communication overhead, often by an order of magnitude, compared to communication-heavy strategies like reactive CBBA (\texttt{cbba}) and even state-of-the-art methods like Clustering-CBBA (\texttt{c-cbba}). For instance, under high task saturation (V500), it used 34x fewer messages than \texttt{cbba} and 10x fewer than \texttt{c-cbba}. Crucially, this significant reduction in network traffic is achieved while maintaining top-tier mission effectiveness, delivering task completion rates statistically indistinguishable from the best-performing (but less efficient) periodic baseline (\texttt{cbba-tree}) across various scenarios.

This advantageous balance emerges from the framework's event-triggered design philosophy. By activating network consensus selectively based on local assessments of information value ($Tri_{\Delta bid}$) or potential conflict ($Tri_{conflict}$), and by allowing the swarm to adapt its fallback communication frequency ($I_{adapt}$) based on environmental dynamics, CBBA-ETC minimizes unnecessary network interactions. This results in a collective system that intelligently self-regulates its network usage, conserving bandwidth and energy, which is critical for practical deployments in constrained environments. Furthermore, the framework demonstrated significant resilience, maintaining high relative performance despite communication degradation (up to 30\% packet loss) and robustness to both transient action execution failures and permanent agent loss, highlighting the effectiveness of the integrated architecture for complex, unpredictable scenarios.

The CBBA-ETC architecture serves as a practical blueprint for designing adaptive and resource-efficient networked robotic systems. The principle of combining event-triggered communication logic with decentralized consensus and robust execution is applicable beyond SAR to other domains requiring efficient coordination under network constraints, such as logistics, environmental monitoring, and precision agriculture.

Despite the promising results, limitations exist. The evaluation relied on simulation, abstracting real-world network complexities. While resilient, the reliance on fewer critical messages implies potential sensitivity to their loss, although performance remained high under tested packet loss conditions . Finally, the event-triggering rules employ empirically tuned static parameters, suggesting an opportunity for further optimization .

\subsection{Future Work Directions}

Future efforts will focus on enhancing and validating the framework, particularly concerning its network interactions. Validating CBBA-ETC on physical robot platforms under realistic network conditions (e.g., variable latency, limited bandwidth) is a crucial next step. We plan to investigate machine learning techniques for agents to autonomously learn optimal communication triggering thresholds and adaptive interval policies, potentially adapting to real-time network quality metrics. Exploring the framework's performance over challenging network topologies (sparse, intermittent) and integrating network-aware routing or utility functions are also key directions. Enhancing the robustness of the underlying execution model (BTs) specifically against communication delays or temporary network partitioning could further improve overall system resilience. Finally, integrating high-level reasoning capabilities, perhaps using LLMs, to translate mission objectives into context-aware communication strategies remains an interesting avenue for improving human-swarm interaction over networks.

% \appendix
% \section{My Appendix}
% Appendix sections are coded under \verb+\appendix+.

\bibliographystyle{cas-model2-names}

% \bibliography{main}

\end{document}